\documentclass[12pt]{article}
\usepackage{amsmath, amssymb, bbm}
\usepackage{graphicx,psfrag,epsf}
\usepackage{enumerate}
\usepackage{natbib}
\usepackage{mdwlist}
\usepackage{url} 
\usepackage[ruled,vlined]{algorithm2e}
\usepackage[letterpaper,margin=1in]{geometry}
\usepackage{titlesec}
\usepackage{blkarray}
\usepackage{multirow}
\usepackage{bigdelim}
\usepackage[usenames,dvipsnames]{xcolor} 
\usepackage{threeparttable}

\titleformat*{\section}{\large\bfseries}
\titleformat*{\subsection}{\normalsize\bfseries}
\titleformat*{\subsubsection}{\normalsize\bfseries}

\newcommand{\blind}{1}

\newcommand{\bx}{\boldsymbol{x}}
\newcommand{\beff}{\boldsymbol{f}}
\newcommand{\bu}{\boldsymbol{u}}
\newcommand{\bt}{\boldsymbol{t}}
\newcommand{\by}{\boldsymbol{y}}
\newcommand{\bd}{\boldsymbol{d}}
\newcommand{\bI}{\boldsymbol{I}}
\newcommand{\bm}{\boldsymbol{m}}
\newcommand{\bM}{\boldsymbol{M}}
\newcommand{\bC}{\boldsymbol{C}}
\newcommand{\ba}{\boldsymbol{a}}
\newcommand{\bA}{\boldsymbol{A}}
\newcommand{\bF}{\boldsymbol{F}}
\newcommand{\bB}{\boldsymbol{B}}

\newcommand{\be}{\boldsymbol{e}}

\newcommand{\E}{\operatorname{E}}
\newcommand{\med}{\operatorname{Med}}
\newcommand{\var}{\operatorname{var}}

\newcommand{\argmax}{\operatorname*{arg\,max}}
\newcommand{\tr}{\operatorname{tr}}
\newcommand{\der}{\mathrm{d}}

\newcommand{\mX}{\mathcal{X}}
\newcommand{\mY}{\mathcal{Y}}
\newcommand{\mD}{\mathcal{D}}
\newcommand{\mT}{\mathcal{T}}
\newcommand{\mM}{\mathcal{M}}

\newcommand{\btheta}{\boldsymbol{\theta}}

\newcommand{\bgamma}{\boldsymbol{\gamma}}
\newcommand{\btau}{\boldsymbol{\tau}}
\newcommand{\bLambda}{\boldsymbol{\Lambda}}
\newcommand{\bdelta}{\boldsymbol{\delta}}

\newcommand{\bone}{\boldsymbol{1}}

\newcommand{\T}{{\rm{T}}}

\begin{document}

\def\spacingset#1{\renewcommand{\baselinestretch}%
{#1}\small\normalsize} \spacingset{1}


\if1\blind
{
  \title{\bf \large Bayesian optimal design for ordinary differential equation models with application in biological science}
  \author{\normalsize Antony M. Overstall\thanks{
   CONTACT Antony M. Overstall; A.M.Overstall@southampton.ac.uk; Southampton Statistical Sciences Research Institute, University of Southampton, Southampton, SO17 1BJ, UK.} $^\dagger$, David C. Woods$^\dagger$ and Ben M. Parker$^\ddagger$ \hspace{.2cm}\\
    \normalsize $^\dagger$Southampton Statistical Sciences Research Institute, \\ \normalsize University of Southampton, United Kingdom\\
    \normalsize $^\ddagger$School of Computing and Engineering,\\ \normalsize University of West London, United Kingdom\\}
    \date{}
  \maketitle
} \fi

\if0\blind
{
  \bigskip
  \bigskip
  \bigskip
  \begin{center}
    {\bf \large Bayesian optimal design for ordinary differential equation models with application in biological science}
\end{center}
  \medskip
} \fi

\bigskip
\begin{abstract}

Bayesian optimal design is considered for experiments where the response distribution depends on the solution to a system of non-linear ordinary differential equations. The motivation is an experiment to estimate parameters in the equations governing the transport of amino acids through cell membranes in human placentas. Decision-theoretic Bayesian design of experiments for such nonlinear models is conceptually very attractive, allowing the formal incorporation of prior knowledge to overcome the parameter dependence of frequentist design and being less reliant on asymptotic approximations. However, the necessary approximation and maximization of the, typically analytically intractable, expected utility results in a computationally challenging problem. These issues are further exacerbated if the solution to the differential equations is not available in closed-form. This paper proposes a new combination of a probabilistic solution to the equations embedded within a Monte Carlo approximation to the expected utility with cyclic descent of a smooth approximation to find the optimal design. A novel precomputation algorithm reduces the computational burden, making the search for an optimal design feasible for bigger problems. The methods are demonstrated by finding new designs for a number of common models derived from differential equations, and by providing optimal designs for the placenta experiment.     

\end{abstract}

\noindent%
\vfill

\newpage
\spacingset{1.45} 
\section{Introduction \label{sec:INTRO}}

The dynamics behind a complex physical process can often be described by a set of non-linear ordinary differential equations, where the solution to these equations represents the evolution of system states with respect to time. It is common for the system of equations to depend on some unknown physical properties (parameters) of the process in question and, potentially, on some additional controllable variables. In this paper, new methods are presented for designing experiments for the estimation of statistical models built on the solution to such a system of equations; that is, choosing the most informative combinations of time points and values of the controllable (design) variables at which observations of the physical process should be made. A decision-theoretic approach is adopted, and hence the quality of a design is measured via the expectation of a utility function chosen to encapsulate the aims of the experiment.

We assume equations with $s$ system states $\bu(t;\,\bx,\btheta) = [u_{1}(t; \bx,\btheta), \ldots, u_{s}(t;\,\bx,\btheta)]^\T$ modeled as a function of time $t$ and $v$ design variables with values held in the treatment vector $\bx\in\mX\subset\mathbb{R}^v$. The $p$-vector $\btheta\in\Theta\subset\mathbb{R}^p$ holds the physical parameters requiring estimation. For notational simplicity, the dependence of the system states on $\bx$ and $\btheta$ is usually suppressed, with $\bu(t) = \bu(t;\,\bx,\btheta)$, unless multiple treatments or parameter vectors are being considered. We mostly find designs for initial value problems, with $\bu(t)$ defined via equations
\begin{equation}
\left\{
\begin{array}{lll}
\dot{\bu}(t) & = & \beff\left(\bu(t),t,\bx;\,\btheta\right)\quad \mbox{for $t \in \mT = \left[T_0,T_1\right]$; } 0 \le T_0 < T_1\,,\\
\bu(T_0) & = & \bu_{0}\,,
\end{array}\right.
\label{eq:odes1}
\end{equation}
where $\dot{\bu}(t)$ is the gradient vector of $\bu(t)$ with respect to time $t$, $\bu_{0} = (u_{01},\ldots,u_{0s})^\T \in \mathbb{R}^s$ denotes initial conditions and, for given $\btheta$, $\beff :\mathbb{R}^s \times \mT \times \mX \to \mathbb{R}^s$ is a continuous function satisfying the Lipschitz condition \citep[see][p. 3]{I2009}. This latter assumption ensures equation~(\ref{eq:odes1}) has a unique solution.

Our research is motivated by experiments to study the transport of serine, an amino acid, within a human placenta. Specifically, interest is in the movement of serine across a placental cell membrane (called a vesicle). In the experiments, initial amounts ($\mu$l) of both radioactive and non-radioactive serine are placed exterior and interior to the vesicle, and then the amount of radioactive serine interior to the vesicle is measured at a series of time points. The experimenters have control over initial amounts of both the interior and exterior non-radioactive serine for each experiment, and the times (in seconds) at which observations are taken. The theoretical interior amounts of radioactive and non-radioactive serine at time $t$ form the $s=2$ system states, $\bu(t) = [u_{1}(t), u_{2}(t)]^\T$, with the $v=2$ design variables, $\bx = (x_{1},x_{2})^\T\in [0, 1000]^2$, being, respectively, the exterior amounts of radioactive and non-radioactive serine at time $t=0$. The equations governing the evolution of the system states are
\begin{equation}
\left. \begin{array}{lcl}
\dot{u}_{1}(t) & = & \frac{x_{1}\left(u_{2}(t) + \theta_2\theta_4\right) - u_{1}(t)\left(x_2 + \theta_2\theta_3\right)}{u^\star\left(\bu(t),t,\btheta,\bx\right)}\,,\\
\dot{u}_2(t) & = & \frac{x_2\left(u_1(t) + \theta_2\theta_4\right) - u_2(t)\left(x_1 + \theta_2\theta_3\right)}{u^\star\left(\bu(t),t,\btheta,\bx\right)}\,,\\
u_1(0) & = & u_{01}\,,\\
u_2(0) & = & u_{02}\,,
\end{array} \right\} t \in [0,600],
\label{eq:odes2b}
\end{equation}
where
$$u^\star\left(\bu(t),t,\btheta,\bx\right) = \frac{1}{\theta_1}\left\{2x^\star_{12}u^\star_{12}(t) + (1+\theta_2)\left[\theta_4x^\star_{12} + \theta_3u^\star_{12}(t)\right] + 2\theta_3\theta_4\right\}\,,$$
$u^\star_{12}(t) = u_1(t) + u_2(t)$, $x^\star_{12} = x_1 + x_2$, and initial conditions $\bu_0 = \left(u_{01},u_{02}\right)^{\T} \in [0,1000]^2$ are the amounts of radioactive and non-radioactive serine interior to the vesicle at time $t=0$. Here, the four physical parameters correspond to the maximum uptake ($\theta_1$), the proportion of the reaction occurring through active transport ($\theta_2$) and two reaction rates ($\theta_3$ and $\theta_4$). The values of these parameters are of scientific interest. See \citet{Panitchob_et_al2015} and \citet{Widdows_et_al2017} for further details of the model and experiment.


To model experimental data from a physical process governed by~\eqref{eq:odes1}, we build a statistical model linking the physical parameters to noisy observations of the system states, or functions thereof, via an assumed data-generating process dependent on the solution to the equations (see, for example, \citealt{RHCC2007}). We also assume that an experiment can be conducted where these observations are collected at various different times and, possibly, from multiple runs of the experiment with different combinations of values of the design variables. Let $n$ denote the number of runs in the experiment, with the $j$th run being made for treatment $\bx_j = (x_{1j},\ldots,x_{vj})^\T$ and initial conditions $\bu_{0j} = (u_{j01},\ldots,u_{j0s})^\T$, with observations being made at time points $\bt_j = (t_{j1},\ldots,t_{jn_j})^\T$ ($j=1,\ldots,n$). At each time point, observations $\by_{jl} \in \mY_{jl}\subset\mathbb{R}^c$ are taken on $c\le s$ different responses. Let $\by_j^\T = (\by_{j1}^\T,\ldots,\by_{jn_j}^\T)$ be the $cn_j$-vector of observations from the $j$th run, and $\by = \left(\by_1^\T,\ldots,\by_n^\T\right)^\T$ be the vector of observations from the whole experiment.

We describe the experimental data $\by$ using statistical model
\begin{equation}
\by | \btheta, \bgamma, \bd \sim \mathrm{F}\left(\btheta, \bgamma;\,\bd\right),
\label{eq:odes2}
\end{equation}
with $\mathrm{F}$ a specified probability distribution, $\bgamma \in \Gamma\subset\mathbb{R}^q$ a $q$-vector of nuisance parameters, and $\bd \in \mD$ a vector specifying the design, chosen from the space of possible designs $\mD$. 
The dependence of~(\ref{eq:odes2}) on physical parameters $\btheta$ and design $\bd$ is through the solution to equations~(\ref{eq:odes1}). The most common form of this dependence, assumed in this paper, is via the expected response,
$$\E\left(\by_{jl} | \btheta,\bx_j,t_{jl}\right) = g\left(\bu(t_{jl}),\btheta\right)\,,$$
with $g:\mathbb{R}^s \times \Theta \to \mY_{jl}$ an assumed function. However, the methodology developed here is also immediately applicable to other types of dependency.   

Here, we find designs for experiments where one or more of the treatments $\bx_1,\ldots,\bx_n$, observation times $t_{j1},\ldots,t_{jn_j}$, for $j=1,\dots,n$, and initial conditions $\bu_{01},\ldots,\bu_{0n}$ are under the experimenters' control. In practice, some of these may be fixed by the protocol of the experiment. We also find designs where the initial conditions are unknown, and included in the vector of parameters.

In the human placenta experiment, the initial quantities of non-radioactive serine interior ($u_{02}$) and exterior ($x_2$) to the vesicle can be varied, with the initial quantities of radioactive serine ($u_{01}$ and $x_1$) fixed by the experimental protocol. The $c=1$ observed response, $y_{jl}$, is the amount of interior radioactive serine at time $t_{jl}$ $(j=1,\ldots,n;\,l=1,\ldots,n_j)$. A statistical model is assumed where $\E\left(y_{jl}|\btheta,\bx_j,t_{jl}\right) = u_1(t_{jl};\,\bx_j,\btheta)$.
Hence, for this experiment $g(\bu,\btheta) = u_1$. The design consists of $n$ combinations of initial quantities of exterior and interior non-radioactive serine, $x_{2j}$ and $u_{02j}$, along with corresponding observation times $t_{j1},\dots,t_{jn_j}$; that is, $\bd = [(x_{21},u_{021})^\T,\ldots,(x_{2n}, u_{02n})^\T,\bt_1^\T,\ldots,\bt_n^\T]^\T$.

Previous research on optimal design for models formed as the solution of ordinary differential equations has focussed on frequentist methods for models with additive normally distributed errors, with a design selected that maximizes a function of the Fisher information matrix for $\btheta$ (e.g. \citealp{AB2002} and \citealp{RS2014}). The inverse of the Fisher information matrix provides an asymptotic approximation to the variance-covariance matrix for maximum likelihood estimators of $\btheta$. As is usual for nonlinear models, the information matrix depends on the value of $\btheta$, which is uncertain prior to the experiment. The most common methodology to overcome this dependence is the adoption of pseudo-Bayesian techniques, where a design is found that maximizes the expectation of the function of the information matrix with respect to a prior distribution for $\btheta$. Numerical methods are used to obtain the derivatives of the expected response with respect to $\btheta$ that are necessary to obtain the information matrix. Most commonly, the ``direct method'' \citep{VV1984} is employed, with an additional set of differential equations being defined that then also require numerical solution. Many developments in this area have occurred in the chemical engineering literature, labeled ``model-based design of experiments''; see \citet{FM2008} for a review.   

In contrast to the above approaches, in this paper, we present and apply the first methods for decision-theoretic Bayesian optimal design for models formed from ordinary differential equations. Although straightforward in principle, Bayesian optimal design faces a number of practical difficulties. Firstly, assessment of a given design requires evaluation of an expected utility depending on high-dimensional and typically intractable integrals. Secondly, maximization of the expected utility presents a high-dimensional and stochastic optimization problem. See \citet{RDMP2016} and \citet{WOAW2017} for recent reviews. 

To address the high-dimensional optimization problem, we extend and apply the approximate coordinate exchange (ACE) algorithm recently proposed by \citet{OW2015}. A brief description of the algorithm is provided in Section~\ref{EX_PRELIM} and Appendix~\ref{app:ACEalg}.

The computational burden of optimal Bayesian design is exacerbated when the model evaluations (systems states) are only available as the numerical solution to the differential equations. In addition to increasing the computational expense of evaluating the expected utility, numerical solutions introduce an additional source of uncertainty through the numerical errors that result from finite discretization of the time interval $\mT$. We evaluate the expected utility by embedding within a Monte Carlo approximation scheme an adaption of the probabilistic solution to systems of differential equations proposed by \citet{CCGC2015}; see Section~\ref{sec:probsol}. In essence, this approach accounts for uncertainty due to discretization error by placing a joint Gaussian process prior on both the system states and time derivatives, and predicts future system states by conditioning on the derivatives. In Section~\ref{sec:ODEdesign}, after introducing the foundations of Bayesian design, we propose innovative precomputation of variance and covariance quantities that substantially reduces the computational burden of incorporating the probabilistic solution into a Bayesian design strategy. Our approach makes it possible to search for multi-variable designs which would otherwise be computationally infeasible.

We demonstrate the effectiveness for optimal design of the combination of Monte Carlo approximation, probabilistic numerics and cyclic descent for a variety of exemplar models in Section~\ref{sec:EXAMPLES}. The differing complexities of the problems addressed showcase the flexibility of the methodology. In Section~\ref{sec:EX_PLAC} we apply the methodology to a realistic statistical model for the human placenta example, based on the solution to~\eqref{eq:odes2b}, and compare to designs proposed by the experimenters. We find designs for the goals of parameter estimation and model selection, where the aim is to  determine if a simpler model with $\theta_3=\theta_4$ (i.e. the two reaction rates equal) is an adequate description for the data.

\section{Probabilistic solutions to ordinary differential equations}\label{sec:probsol}
When working with numerical models implemented via computer code, it has become standard to build statistical approximations, or emulators, by performing a computer experiment to obtain model outputs at carefully selected input combinations. Most commonly, a Gaussian process (GP) prior is assumed to describe the output from the model, with the emulator formed from the updated posterior GP (conditioned on the model output from the computer experiment); see \citet{sacks} and \citet{SWN2003}. In contrast, central to the \citet{CCGC2015} methodology is the adoption of a GP prior for the $h$th derivative function $\dot{u}_h(\cdot)$, $h=1,\ldots,s$, defined via mean and covariance functions $\dot{m}_{0h}(\cdot)$ and $\dot{C}_0(\cdot, \cdot)$, where we assume a common covariance function for each of the $s$ derivatives. Such a prior implies that for any finite collection of times $\bt = (t_1,\ldots, t_w)^{\T}$, the joint distribution of $\dot{u}_h(\bt) = [\dot{u}_h(t_1),\ldots,\dot{u}_h(t_w)]^{\T}$ will be multivariate normal $N(\dot{\bm}_{0h}(\bt), \dot{\bC}_0(\bt, \bt))$, with $n$-vector $\dot{\bm}_{0h}(\bt)$ having $l$th entry $\dot{m}_{0h}(t_l)$ and, for vector $\bt^\prime = (t^\prime_1,\ldots,t^\prime_{w^\prime})^\T$, $w\times w^\prime$ matrix $\dot{\bC}_0(\bt, \bt^\prime)$ having $lk$th entry $\dot{C}_0(t_l, t^\prime_k)$. A joint Gaussian process prior for both $\dot{u}_h(\cdot)$ and the solution function $u_h(\cdot)$ then follows directly, implying  the joint distribution
$$
\left(\begin{array}{c}
\dot{u}_h(\bt)\\
u_h(\bt^\prime) \end{array}\right) \sim N\left( \left(
\begin{array}{c}
\dot{\bm}_{0h}(\bt)\\
\bm_{0h}(\bt^\prime) \end{array}\right),\left(
\begin{array}{cc}
\dot{\bC}_0(\bt, \bt) & \bar{\bC}_0(\bt,\bt^\prime) \\
\bar{\bC}_0(\bt^\prime,\bt) & \bC_0(\bt^\prime,\bt^\prime)
\end{array}\right) \right)\,,
$$
with $w$-vector $\bm_{0h}(\bt)$ having $l$th entry $m_{0h}(t_l) = \int_0^{t_l}\dot{m}_{0h}(z)\,\mathrm{d}z + u_{0h}$, $w\times w^\prime$ matrix $\bC_0(\bt,\bt^\prime)$ having $lk$th entry $C_0(t_l,t^\prime_k) = \int_0^{t_l}\int_0^{t_k^\prime}\dot{C}_0(z,z^\prime)\,\mathrm{d}z\mathrm{d}z^\prime$, and $w\times w^\prime$ cross-covariance matrix $\bar{\bC}_0(\bt,\bt^\prime)$ having $lk$th entry $\bar{C}_0(t_l,t^\prime_k)= \int_0^{t_k^\prime}\dot{C}_0(t_l,z)\,\mathrm{d}z$; see also \citet{SMLLR2003} and \citet{HSLHHHA2013}. Hence, solution vector $u_h(\bt) = [u_h(t_1),\ldots,u_h(t_n)]^\T$ follows the multivariate normal distribution $N(\bm_{0h}(\bt), \bC_0(\bt, \bt^\prime))$. Note that definition of the covariance function of $u_h(\bt)$ via integration ensures $C_0(0, 0) = 0$ and hence enforces the boundary condition $u_h(0) = u_{0h}$.

\spacingset{1}
\begin{algorithm}
\DontPrintSemicolon

\nl Set $\bLambda_1=0$ and $\beff_1 = \beff(\bu_{0}, T_0, \bx;\,\btheta)$\;
\nl \label{alg:update:iter}\For{$r=1,\ldots,N-1$}{
(a) Set $\btau_r = (\tau_1,\ldots, \tau_r)^\T$ \\[2ex]

(b) Compute \\[1ex]
$\bB_r = (\dot{\bC}_0(\btau_r,\btau_r) + \bLambda_r)^{-1}$ \\[1ex]
$\ba_r = \bB_r\bar{\bC}_0(\btau_r, \tau_{r+1})$ \\[1ex]
$C_r = C_0(\tau_r, \tau_r) - \bar{\bC}_0(\tau_{r+1}\btau_r)\bB_r\bar{\bC}_0(\btau_r, \tau_{r+1})$ \\[1ex]
$\dot{C}_{r+1} = \dot{C}_0(\tau_{r+1}, \tau_{r+1}) - \dot{\bC}_0(\tau_{r+1},\btau_r)\bB_r\dot{\bC}_0(\btau_r, \tau_{r+1})$ \\[1ex]
$\bLambda_{r+1} = \mathrm{diag}\{\bLambda_r, \dot{C}_{r+1}\}$ \\[2ex]

(c) Compute \\[1ex]
$\bm_r = \bu_{0} + \bF_r^{\T}\ba_r$\,, where $\bF_r$ is the $r\times s$ matrix with $k$th row $\beff_k$ $(k=1,\ldots,N-1)$\\[2ex]

(d) Sample \\[1ex]
$\bu(\tau_{r+1}) \sim N(\bm_r, C_r\bI_{S})$ \\[1ex]
and compute \\[1ex]
$\beff_{r+1} =  \beff(\bu(\tau_{r+1}), \tau_{r+1}, \bx;\, \btheta)$ \\[2ex]}
\nl \label{alg:update:BN} Compute \\[1ex]
$\bB_N = (\dot{\bC}_0(\btau_N, \btau_N) + \bLambda_N)^{-1}$ \\[1ex]
$\bA_N(\bt) = \bB_N\bar{\bC}_0(\btau,\bt)$ \\[1ex] 
$\bM_N(\bt) = \bone_n\otimes\bu_{0}^{\T} + \bA_N^\T(\bt)\bF_N$, with $\bone_n$ the $n$-vector with all entries equal to one and $\bF_N$ the $N\times s$ matrix with $k$th row $\beff_k$ $(k=1,\ldots,N)$ \\[1ex]
$\bC_N(\bt,\bt) = \bC_0(\bt) - \bar{\bC}_0(\bt,\btau) \bB_N\bar{\bC}_0(\btau,\bt)$ \\[2ex]
\nl For $h=1,\ldots,s$, sample \\[1ex]
$u_h(t_1),\ldots, u_h(t_n) \sim N\left(\bM_N(t)\be_h, \bC_N(\bt,\bt)\right)$, where $\be_h$ is the $h$th unit vector

\caption{Sequential updating and sampling for time points $\bt = (t_1,\ldots,t_w)^{\T}$ of the joint Gaussian process for the derivative and solution for the $s$ system states for initial values $\bu_0$, treatment vector $\bx$, physical parameters $\btheta$ and evaluation grid $\btau = (\tau_1, \ldots, \tau_N)^{\T}$, with $\tau_1=T_0$. (Adapted from \citealp{CCGC2015}). \label{alg:update}}
\end{algorithm}
\spacingset{1.45}

For a given $\bx$ and $\btheta$, this prior distribution can be updated using derivative evaluations on a grid $\btau = (\tau_1,\ldots,\tau_N)^\T$ of time points via Algorithm~\ref{alg:update} by sequentially conditioning on $f(\bu,\tau_{r+1},\bx;\,\btheta)$ calculated for solution state $u_h$ sampled from the posterior distribution at point $\tau_r$. The final marginal Gaussian process for $u_h(t)$ has mean and covariance functions given by
$$
m_{Nh}(t) = u_{0h} +  \bar{\bC}_0(t,\btau)\bB_N\bF_N\be_h\,,\quad C_N(t,t^\prime) = C_0(t,t^\prime) - \bar{\bC}_0(t,\btau) \bB_N\bar{\bC}_0(\btau,t^\prime)\,,
$$
for $h=1,\ldots,s$, where $\be_h$ is the $h$th unit vector, and the $N\times s$ matrix of derivative evaluations $\bF_N$ and the updated $N\times N$ derivative covariance matrix $\bB_N$ are defined as in Algorithm~\ref{alg:update}.

\begin{figure}
\centering
\includegraphics[scale=.9, viewport = 45 505 548 671, clip=true]{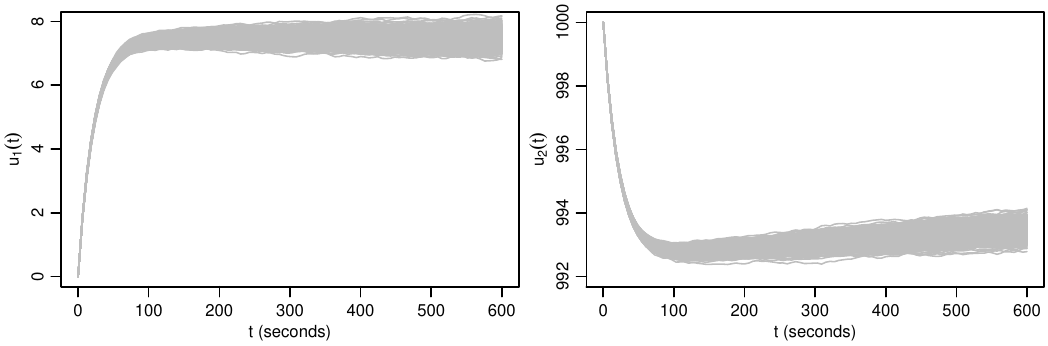}
\caption{\label{fig:plac_sols}Plots showing 1000 draws from the probabilistic solution of $u_1(t)$ and $u_2(t)$ against $t$ for system of equations~\eqref{eq:odes2b} that describe the transport of serine in a human placenta.}
\end{figure}

\citet{CCGC2015} allowed covariance function $\dot{C}_0(t,t^\prime)$ to depend on hyperparameters controlling the scale and length of the covariances. Given experimental data, a joint posterior distribution for the model parameters and hyperparameters can be sampled by embedding the probabilistic solution to the differential equations within a Markov chain Monte Carlo scheme. \citet{CCGC2015} also suggested possible fixed values for the covariance hyperparameters. In Section~\ref{sec:extode} we demonstrate the computational savings that can be achieved for optimal design via precomputing of various posterior quantities when these parameters are fixed. 

Figure~\ref{fig:plac_sols} presents 1000 draws from probabilistic solutions for the placenta example following equations~\eqref{eq:odes2b}. Updated Gaussian processes for $u_1(t)$ and $u_2(t)$ were generated using Algorithm~\ref{alg:update} assuming a squared exponential covariance function for $\dot{C}(t,t^\prime)$ (see \citealp{RW2006}, p.~83). An evaluation grid $\btau$ with $N=501$ evenly spaced time points was used, and the solution sampled for time $t\in [T_0, T_1] = [0, 600]$ seconds with physical parameters $\btheta = (200, 0.05, 100, 100)^\T$, initial values $\bu_0 = (0, 1000)^\T$ and treatment $\bx = (7.5, 1000)^\T$. Note how the uncertainty in the solution increases as $t$ increases away from $t=T_0=0$ where we know, in this example, the true value of $\bu(t)$.

\section{Bayesian design for ordinary differential equation models}
\label{sec:ODEdesign}

\subsection{Decision-theoretic Bayesian optimal design \label{sec:BOD_DTA}}

Design of experiments fits naturally within a Bayesian framework, with the decision on what design $\bd$ to employ made before the data is collected. Hence it is natural to use available prior information to inform this choice. This information includes the form of statistical model~\eqref{eq:odes2} including any underpinning physical theory, for example, as encapsulated in equations such as~\eqref{eq:odes1}. It also includes any prior information on the values of the parameters $\btheta,\bgamma$, captured via a prior density $\pi(\btheta,\bgamma)$, which we assume is independent of the design. 

A decision-theoretic Bayesian optimal design, $\bd^\star$, maximizes the expectation of a specified utility function $\phi(\btheta,\by,\bd)$ with respect to the unknowns prior to experimentation,
\begin{align*}
\Phi(\bd^\star) & = \max_{\bd\in\mD} \E\left[\phi(\btheta,\by,\bd) | \bd\right] \nonumber\\
 & = \max_{\bd\in\mD} \int_{\Theta. \mY} \phi(\btheta,\by,\bd)\pi(\btheta,\by | \bd)\,\der\btheta\,\der\by\,,
\end{align*}
where the joint distribution of the unknown physical parameters and responses, conditional on the design used for data collection, can be decomposed as 
$$\pi(\btheta,\by | \bd) = \int_\Gamma\pi(\by | \btheta, \bgamma, \bd)\pi(\btheta,\bgamma)\,\der\bgamma\,,$$ 
and hence, when regarded as a function of $\btheta$ alone, is proportional to the posterior density. See the seminal review paper by \citet{CV1995}.       

The function $\phi(\btheta,\by,\bd)$ quantifies the utility, relative to the aims of the experiment, for choosing design $\bd$ when we obtain data $\by$ under physical parameters $\btheta$. Its choice should reflect the goals of the experiment. Here, we apply the following exemplar utility functions:
\begin{enumerate}
\item Negative squared error loss (NSEL) for estimation of $\btheta$:
$$
\phi(\btheta,\by,\bd) = -\lVert\btheta - \E(\btheta | \by,\bd)\rVert_2^2\,,
$$
with $\lVert\cdot\rVert_p$ denoting the $l_p$-norm and $\E(\btheta | \by,\bd)$ the posterior mean, where expectation is taken with respect to the marginal density $\pi(\btheta | \by,\bd) = \int_\Gamma\pi(\btheta,\bgamma | \by,\bd)\,\der\bgamma$. It can be shown that the expected utility simplifies to
\begin{align*}
\Phi(\bd) & = -\int_\mY \tr\left\{\var(\btheta|\by,\bd)\right\}\pi(\by|\bd)\,\der\by\,,
\end{align*}
the negative expected value of the posterior variance-covariance matrix for $\btheta$ with respect to the marginal distribution of the response.
\item Negative absolute error loss (NAEL) for estimation of $\btheta$:
$$
\phi(\btheta,\by,\bd) = -\lVert\btheta - \med(\btheta | \by,\bd)\rVert_1\,,
$$
with $\med(\btheta | \by, \bd)$ the vector of marginal posterior medians of the physical parameters.
\item Shannon information gain (SIG) for $\btheta$:
\begin{equation}\label{eq:sil}
\phi(\btheta,\by,\bd) = \log\pi(\by | \btheta,\bd) - \log\pi(\by | \bd)\,,
\end{equation}
where
$$ 
\pi(\by|\bd) = \int_\Theta\pi(\by | \btheta, \bd)\pi(\btheta)\,\der\btheta\,,\qquad \pi(\by | \btheta, \bd) = \int_\Gamma \pi(\by | \btheta,\bgamma, \bd)\pi(\bgamma)\,\der\bgamma\,.
$$
Maximizing the expectation of~\eqref{eq:sil} is equivalent to maximizing the expected Kullback-Liebler divergence between the prior and posterior distributions \citep{CV1995}.
\suspend{enumerate}
For the human placenta example, we also employ two bespoke utility functions tailored to the problems of point estimation and model selection.

\resume{enumerate}
\item 0-1 utility for estimation of $\btheta$:
$$
\phi(\btheta, \by, \bd) = \mathbbm{1}_{\check{\Theta}}\left[\E(\btheta | \by,\bd)\right]\,,
$$
with $\mathbbm{1}_{\check{\Theta}}$ the indicator function for the product set 
$$
\check{\Theta} = \prod_{i=1}^p \check{\Theta}_i = \left\{(\check{\theta}_1,\ldots,\check{\theta}_p)\,\vert\,\check{\theta}_i \in \check{\Theta}_i\, \forall i \in \{1,\ldots,p\}\right\}\,,
$$ 
where $\check{\Theta}_i = \left\{\check{\theta}\,\vert\, \theta_i - \delta_i \le \check{\theta} \le \theta_i + \delta_i\right\}$, and $\bdelta = (\delta_1,\ldots,\delta_p)^\T$ is a specified tolerance vector. That is, the utility is equal to 1 if, for all $i=1,\ldots,p$, the $i$th element of the posterior mean vector $\E(\btheta | \by,\bd)$ lies within $\delta_i$ of the corresponding element of $\btheta$.
\suspend{enumerate}  

For the final utility function considered we redefine the utility as a function of the chosen model $m\in\mM$.

\resume{enumerate}
\item 0-1 utility for model selection: 
$$
\phi(m, \by, \bd) = \mathbbm{1}_{m}(m^\star) \,,
$$
where $\mathbbm{1}_{m}$ is the indicator function for the singleton set with element $m$ and $m^\star\in \argmax_{m\in\mM} \pi(m | \by)$ is the model with maximum posterior probability. For this utility, the expected utility is given by
$$
\Phi(\bd) = \sum_{m\in\mM} \pi(m)\int_\mY \phi(m,\by,\bd)\pi(\by | m,\bd)\,\der\by\,,
$$
with $\pi(\by | m,\bd) = \int_{\Theta^{(m)}}\int_{\Gamma^{(m)}}\pi(\by | \btheta^{(m)}, \bgamma^{(m)}, m, \bd)\pi(\btheta^{(m)}, \bgamma^{(m)} | m)\,\der\bgamma^{(m)}\,\der\btheta^{(m)}$, and $\btheta^{(m)}\in\Theta^{(m)}$ and $\bgamma^{(m)}\in\Gamma^{(m)}$ physical and nuisance parameters, respectively, for model $m$.
\end{enumerate}

A barrier to the application of Bayesian design for most nonlinear models, including those considered in this paper, is the analytic intractability of both the utility function (which typically depends on posterior quantities) and expected utility. Numerical methods are therefore required, with a double-loop Monte Carlo approximation being commonly employed \citep{Ryan2003}. Such an approach uses an ``inner'' Monte Carlo sample of size $\tilde{B}$ to approximate any necessary posterior quantities, and then an ``outer'' Monte Carlo sample of size $B$ to approximate the expected utility with respect to the joint distribution of $\by$ and $\btheta$; see also \citet{OW2015}.

We use the approximation
\begin{equation}\label{eq:mc}
\hat{\Phi}(\bd) = \frac{1}{B}\sum_{b=1}^{B} \hat{\phi}(\btheta_b,\by_b,\bd)\,,
\end{equation}
with $\{\btheta_b,\by_b\}_{b=1}^{B}$ a first (outer) sample from the joint distribution of the physical parameters and response, and $\hat{\phi}(\btheta, \by, \bd)$ a further Monte Carlo approximation to the utility function. 

Each of the utility functions above can be approximated using a second (inner) Monte Carlo sample $\left\{\tilde{\btheta}_{\tilde{b}}, \tilde{\bgamma}_{\tilde{b}}\right\}_{\tilde{b}=1}^{\tilde{B}}$ from distribution with density
$\pi(\btheta,\bgamma)$:

\begin{enumerate}
\item NSEL:
$$
\hat{\phi}(\btheta,\by,\bd) = -\lVert\btheta - \hat{\E}(\btheta | \by,\bd)\rVert_2^2\,,
$$
for an importance sampling estimate of $\E(\btheta | \by,\bd)$,
\begin{equation}\label{eq:ispostmean}
\hat{\E}(\btheta | \by,\bd) = \sum_{\tilde{b}=1}^{\tilde{B}}w_{\tilde{b}}\tilde{\btheta}_{\tilde{b}}\,,
\end{equation}
with
\begin{equation}\label{eq:weights}
w_{\tilde{b}} = \frac{\pi(\by | \tilde{\btheta}_{\tilde{b}}, \tilde{\bgamma}_{\tilde{b}}, \bd)}{\sum_{\tilde{b}=1}^{\tilde{B}}\pi(\by | \tilde{\btheta}_{\tilde{b}}, \tilde{\bgamma}_{\tilde{b}},\bd)}\,.
\end{equation}
See \citet{RDMP2016} and references therein.
\item NAEL:
$$
\hat{\phi}(\btheta,\by,\bd) = -\lVert\btheta - \widehat{\med}(\btheta | \by,\bd)\rVert_1\,,
$$
with vector $\hat{\med}(\btheta | \by,\bd)$ having $i$th entry  $\widehat{\med}_i(\btheta | \by,\bd) = (\tilde{\theta}_{i(z)} + \tilde{\theta}_{i(z+1)})/2$ ($i=1,\ldots,p$), where $\tilde{\theta}_{i(1)}\le\cdots\le\tilde{\theta}_{l({\tilde{B}})}$ are the ordered values taken by the $i$th element of the sample  $\left\{\tilde{\btheta}_{\tilde{b}}\right\}_{\tilde{b}=1}^{\tilde{B}}$, $z = \max \{l=1,\ldots, \tilde{B} | \sum_{\tilde{b}=1}^{l}w_{i(\tilde{b})}\le 0.5\}$ and the $w_{i(\tilde{b})}$ are the weights~\eqref{eq:weights} ordered according to $\theta_{i(\tilde{b})}$. 

\item SIG:
$$
\hat{\phi}(\btheta,\by,\bd) = \log\hat{\pi}(\by | \btheta,\bd) - \log\hat{\pi}(\by | \bd)\,,
$$
with
$$
\hat{\pi}(\by | \bd) = \frac{1}{\tilde{B}}\sum_{\tilde{b}=1}^{\tilde{B}}\pi(\by | \tilde{\btheta}_{\tilde{b}}, \tilde{\bgamma}_{\tilde{b}}, \bd)\,,\qquad \hat{\pi}(\by | \btheta, \bd) = \frac{1}{\tilde{B}}\sum_{\tilde{b}=1}^{\tilde{B}}\pi(\by | \btheta, \tilde{\bgamma}_{\tilde{b}}, \bd)\,.
$$

\item 0-1 estimation:
$$
\hat{\phi}(\btheta, \by, \bd) = \mathbbm{1}_{\check{\Theta}}\left[\hat{\E}(\btheta | \by,\bd)\right]\,,
$$
for $\hat{\E}(\btheta | \by,\bd)$ once again the importance sampling estimate~\eqref{eq:ispostmean} of the posterior mean.

\item 0-1 model selection:
$$
\hat{\phi}(m, \by, \bd) = \mathbbm{1}_{m}(\hat{m}^\star)\,,
$$
where $\hat{m}^\star\in \argmax_{m\in\mM}\pi(m)\sum_{\tilde{b}=1}^{\tilde{B}}\pi(\by | \tilde{\btheta}_{\tilde{b}}^{(m)}, \tilde{\bgamma}_{\tilde{b}}^{(m)}, m, \bd)/\tilde{B}$ for $\left\{\tilde{\btheta}^{(m)}_{\tilde{b}}, \tilde{\bgamma}_{\tilde{b}}^{(m)}\right\}_{\tilde{b}=1}^{\tilde{B}}$ a sample from the prior distribution under model $m$ with density $\pi(\btheta^{(m)}, \bgamma^{(m)} | m)$.
\end{enumerate}

The above Monte Carlo approximations $\hat{\phi}$ to the utility functions will introduce some bias into the approximation of the expected utility, as the utilities are nonlinear functions of posterior quantities. In general, this bias will be inversely proportional to the value of $\tilde{B}$, and hence can be made negligible for large inner samples. 


\subsection{Extensions to ordinary differential equation models}\label{sec:extode}

To apply the methodology outlined in the previous section to models built from systems of ordinary differential equations requires incorporation of further steps to account for discretization errors in the numerical solution to the equations, and to mitigate the additional computational cost of multiple evaluations of the numerical solution. The approximations to the expected utilities require repeated sampling of $\by$ from distribution~\eqref{eq:odes2}, and evaluation of the corresponding density function $\pi(\by|\btheta, \bgamma,\bd)$. When the distribution of $\by$ depends on the solution vector, the approximations require at least $B+\tilde{B}$ evaluations of a numerical solution to $\bu(t_{jl};\,\bx_j,\btheta)$ for each $j=1,\ldots, n$ and $l = 1,\ldots,n_j$. In addition to the computational cost of these repeated evaluations, the necessary discretization of the time domain by a numerical solver introduces an additional source of uncertainty that should be accounted for in both the design of the experiment and the subsequent inference.

The probabilistic solution of \citet{CCGC2015}, outlined in Section~\ref{sec:probsol}, fits naturally within a Monte Carlo approximation of the expected utility; for each generated value of the physical parameters $\btheta$, a solution path for $\bu(t)$ is generated from an updated Gaussian process. The uncertainty introduced by the discretization of time is quantified, and updated, via the joint Gaussian process prior for the time derivatives and solution. Algorithm~\ref{alg:approxEU} outlines the steps in generating an approximation to a general utility function $\phi$ using double loop Monte Carlo. As given, Algorithms~\ref{alg:update}--\ref{alg:approxEU} depend on the initial values $\bu_{0j}$ for the $j$th treatment, i.e. the initial values are assumed known. In some situations, learning unknown initial values may be part of the inference problem, i.e. prior distributions are assumed and updated to a posterior distribution in light of the experimental responses. This case can be incorporated into these algorithms by replacing all occurrences of $\bu_{0j}$ by a value $\bu_{0jb}$ generated from the prior distribution in Algorithm~\ref{alg:approxEU}, in an analogy to how the physical parameters $\btheta$ are handled.

\spacingset{1}
\begin{algorithm}
\DontPrintSemicolon
\nl\label{alg:approxEU:inner} \For{$\tilde{b} = 1, \ldots, \tilde{B}$}{
Sample $(\tilde{\btheta}^\T_{\tilde{b}}, \tilde{\bgamma}^\T_{\tilde{b}})^\T\sim\pi(\btheta,\bgamma)$ (the prior distribution)\\[1ex]
\For{$j = 1, \ldots, n$}{
\For{$l = 1, \ldots, n_j$} {
Sample $u_s(t_{jl};\,\bx_j,\tilde{\btheta}_{\tilde{b}})$ using Algorithm~\ref{alg:update} \\[1ex]
}
}
}
\nl \For{$b=1,\ldots,B$}{
Sample $(\btheta_b^\T,\bgamma_b^\T)^\T\sim \pi(\btheta,\bgamma)$ (the prior distribution) \\[1ex]
\For{$j = 1, \ldots, n$}{
\For{$l = 1, \ldots, n_j$} {
Sample $u_s(t_{jl};\,\bx_j,\btheta_b)$ using Algorithm~\ref{alg:update}
}
Sample $\by_j |\btheta_b, \bgamma_b, d \sim \mathrm{F}(\btheta_b, \bgamma_b;\,d)$ 
}
Calculate $\hat{\phi}(\btheta_b,\by_b,d)$ using the inner sample generated in step~\ref{alg:approxEU:inner}  \\[1ex]
} 
\nl Calculate $\hat{\Phi}(d) = \frac{1}{B}\sum_{b=1}^B\hat{\phi}(\btheta_b,\by_b,d)$
\caption{Evaluation of the approximate expected utility $\hat{\Phi}(d)$ when the distribution of the response depends on the solution to a ordinary differential equation. \label{alg:approxEU}}
\end{algorithm}
\spacingset{1.45}

Naive implementation of Algorithm~\ref{alg:approxEU} for approximating the expected utility presents a considerable computational challenge, with the matrix computations in steps~\ref{alg:update:iter}(b) and~\ref{alg:update:BN} of Algorithm~\ref{alg:update} being undertaken $\tilde{n}(B+\tilde{B})$ times, with $\tilde{n} = \sum_{j=1}^nn_j$. In particular, calculation of matrix $\bB_N$ requires inversion of an $N\times N$ matrix. This leads to an algorithm with computational complexity $\mathcal{O}(\tilde{n}N^3(B+\tilde{B}))$.

To reduce the computational cost of the algorithm, we can compromise on the choice of covariance function $\dot{C}_0(t,t^\prime)$. Rather than tune the covariance through the selection of different parameter values for each choice of $\bx$ and $\btheta$, we can fix these parameters (e.g. following recommendations in~\citet{CCGC2015}; see Section~\ref{sec:EXAMPLES} for our choices). This allows precomputation of various covariance matrices and vectors, see Algorithm~\ref{alg:precompute}. Such precomputation alleviates the need to invert $\bB_N$ when sampling $u(t)$, reducing the computational complexity of the approximation to $\mathcal{O}(N^3 + \tilde{n}N^2(B+\tilde{B}))$. 

In fact, this precomputation can be performed just once, \textit{prior} to any optimization routine being called. Hence for large experiments and Monte Carlo sample sizes, the computational complexity of the precomputation is essentially fixed, and the complexity of the approximation within the optimization becomes $\mathcal{O}(\tilde{n}N^2(B+\tilde{B}))$. This computational savings makes the optimization feasible for experiment sizes, evaluation grids and Monte Carlo sample sizes for which designs could not otherwise be found.


\spacingset{1}
\begin{algorithm}
\DontPrintSemicolon
\nl Set $\bLambda_1=0$ \; 
\nl \For{$r=1,\ldots,N-1$}{
(a) Set $\btau_r = (\tau_1,\ldots, \tau_r)^\T$ \\[2ex]

(b) Compute \\[1ex]
$\bB_r = (\dot{\bC}_0(\btau_r,\btau_r) + \bLambda_r)^{-1}$ \\[1ex]
$\ba_r = \bB_r\bar{\bC}_0(\btau_r,\tau_{r+1})$ \\[1ex]
$C_r = C_0(\tau_r, \tau_r) - \bar{\bC}_0(\tau_{r+1},\btau)\bB_r\bar{\bC}_0(\btau_r,\tau_{r+1})$ \\[1ex]
$\dot{C}_{r+1} = \dot{C}_0(\tau_{r+1}, \tau_{r+1}) - \dot{\bC}_0(\tau_{r+1},\btau_r)\bB_r\dot{\bC}_0(\btau_r,\tau_{r+1})$ \\[1ex]
$\bLambda_{r+1} = \mathrm{diag}\{\bLambda_r, \dot{C}_{r+1}\}$ \;
}
\nl Compute $\bB_N = \left(\dot{\bC}_0(\btau_N, \btau_N) + \bLambda_N\right)^{-1}$ 
\caption{Precomputation of variances $C_r$, $\dot{C}_{r+1}$, $\bB_r$ and covariances $\ba_r$ for evaluation grid $\btau = (\tau_1,\ldots,\tau_N)^\T$ \label{alg:precompute}}
\end{algorithm}
\spacingset{1.45}

\section{Examples} \label{sec:EXAMPLES}

\subsection{Preliminaries} \label{EX_PRELIM}
In this section we demonstrate the Bayesian design methodology for three common examples of models formed from the solution of ordinary differential equations:
\begin{enumerate}
\item
a compartmental model (Section~\ref{EX_COMP});
\item
a model  formed from the FitzHugh-Nagumo equations (Section~\ref{EX_FITZ});
\item
a model of the JAK-STAT mechanism (Section~\ref{EX_JAKSTAT}).
\end{enumerate}

For each, we use the methodology in Section~\ref{sec:extode} to approximate expected utilities for parameter estimation. Bayesian optimal (or near optimal) designs are found by embedding these Monte Carlo approximations within the ACE algorithm \citep{OW2015}. The ACE algorithm is a cyclic descent, or coordinate exchange, algorithm (see \citealp{MeyerNachtsheim} and \citealp{lange_2013}, p.~171) that performs a sequence of conditional maximizations for each element (coordinate) of $\bd$ in turn, keeping all other elements fixed. Each of these one-dimensional maximizations is performed by constructing a Gaussian process smoother, or emulator, for the Monte Carlo approximation as a function of the coordinate. Use of an emulator alleviates both the computational burden and lack of smoothness associated with the Monte Carlo approximations. This algorithm extends the optimal design via curve fitting methods originally presented by \citet{MP1996} to high-dimensional design problems. The ACE algorithm is outlined in Appendix~\ref{app:ACEalg} and implemented in the \texttt{acebayes} \texttt{R} package (\citealp{OWA2017}, \citealp{ace}), available on \texttt{CRAN}.   

To employ the probabilistic solution to the ordinary differential equations, a choice of covariance function is required for the Gaussian process prior on the derivative functions. The choice of covariance function should be determined by the assumed smoothness of the solutions $u_h(t)$. \citet{CCGC2015} suggested two covariance functions, the squared exponential covariance
\begin{equation}\label{eq:sqexpcov}
\dot{C}_0(t,t^\prime) = \sqrt{\pi}\alpha^{-1}\lambda \exp \left\{ -(t-t^\prime)^2/4\lambda^2\right\}\,,
\end{equation} 
which is infinitely differentiable and hence suitable for smooth solutions, and the piecewise linear uniform covariance
\begin{equation}\label{eq:unifcov}
\dot{C}_0(t,t^\prime) = 
\begin{cases}
\alpha^{-1}\left\{\min(t,t^\prime) - \max(t,t^\prime) + 2\lambda\right\} & \text{for }  \left\{\max(t,t^\prime) - \min(t,t^\prime)\right\}/2 > \lambda\,, \\
0 & \text{otherwise}\,,
\end{cases}
\end{equation}
where $\alpha,\lambda>0$. This latter function is non-differentiable and hence suited to non-smooth solutions. We employ these two functions, with fixed values of $\alpha$ and $\lambda$ to facilitate the precomputation outlined in Section~\ref{sec:extode}. Throughout, we assume the probabilistic solution is calculated on a grid $\btau = (\tau_1,\ldots,\tau_N)^{\T}$ of equally-spaced points and, unless otherwise stated, set $\alpha = N$ and $\lambda = 4(\tau_N-\tau_1)/N$.

The Supplementary Material contains an \texttt{R} package called \texttt{aceodes} and a vignette. The vignette describes how \texttt{aceodes} can be used to reproduce the designs found in the remainder of this section and in Section~\ref{sec:EX_PLAC}.

\subsection{Compartmental model \label{EX_COMP}}
In pharmacokinetics studies, compartmental models are used to describe the distribution of a drug inside a living body. Such models have been routinely used to demonstrate optimal experimental design methodology (see, for example, \citealp{A1993}, \citealp{R2014}, and \citealt{OW2015}). To compare designs found using the probabilistic solution to designs found using an exact solution, we use a simple example where an analytical solution to the differential equations is available. An open one-compartment model is considered with first-order absorption, described by the following system of $s=2$ ordinary differential equations for $t \in [0,24]$ hours:
\begin{equation*}
\begin{array}{lll}
\dot{u}_1(t) & = & - \theta_1 u_1(t)\,, \\
\dot{u}_2(t) & = & (\theta_2/\theta_3) u_1(t) - \theta_2 u_2(t)\,, \\
\bu(0) & = & (D,0)^{\T}\,,
\end{array}
\label{comp_odes}
\end{equation*}
where $u_1(t)$ and $u_2(t)$ are respectively the amounts of drug outside and inside the body, $D$ is the known initial dose, and $\boldsymbol{\theta} = \left(\theta_1,\theta_2,\theta_2\right)^{\T}$ are unknown parameters.

These equations define a homogeneous linear system with constant coefficients, resulting in the analytical solution
\begin{eqnarray}
u_1(t) & = & D \exp \left( - \theta_1 t \right)\,,\nonumber\\
u_2(t) & = & \frac{D\theta_2}{\theta_3(\theta_2-\theta_1)} \left( \exp (-\theta_1t) - \exp (-\theta_2t)\right)\,.\label{eq:compsol2}
\end{eqnarray}
Following \citet{R2014}, we assume $D=400$ and $\log \theta_i \sim \mathrm{N}(\mu_i,0.05)$, independently, for $l=1,2,3$, with $(\mu_1,\mu_2,\mu_3)^{\T} = (\log 0.1, \log 1, \log 20)^{\T}$. The amount of drug inside the body, $y_l$, is observed at observation time $t_l$, and is modeled through assuming $y_l \sim \mathrm{N} \left( u_2(t_l) , \sigma^2 + \tau^2 u_2(t_l)^2 \right)$, independently, where $\sigma^2 = 0.1$ and $\tau^2 = 0.01$. The choice of design here only involves selecting $n=15$ observation times: $t_1,\dots,t_n$. We impose the practically realistic constraint that the observation times have to be at least 15 minutes apart. Such a constraint is straightforward to incorporate into the ACE algorithm (see \citealp{OW2015}).

When applying the probabilistic solution, we assume squared exponential covariance~\eqref{eq:sqexpcov} as the functions $\bu(t)$ are known to be smooth and a discrete evaluation grid, $\boldsymbol{\tau}$, with $N=501$.

\begin{figure}
\begin{center}
\includegraphics[scale=.9]{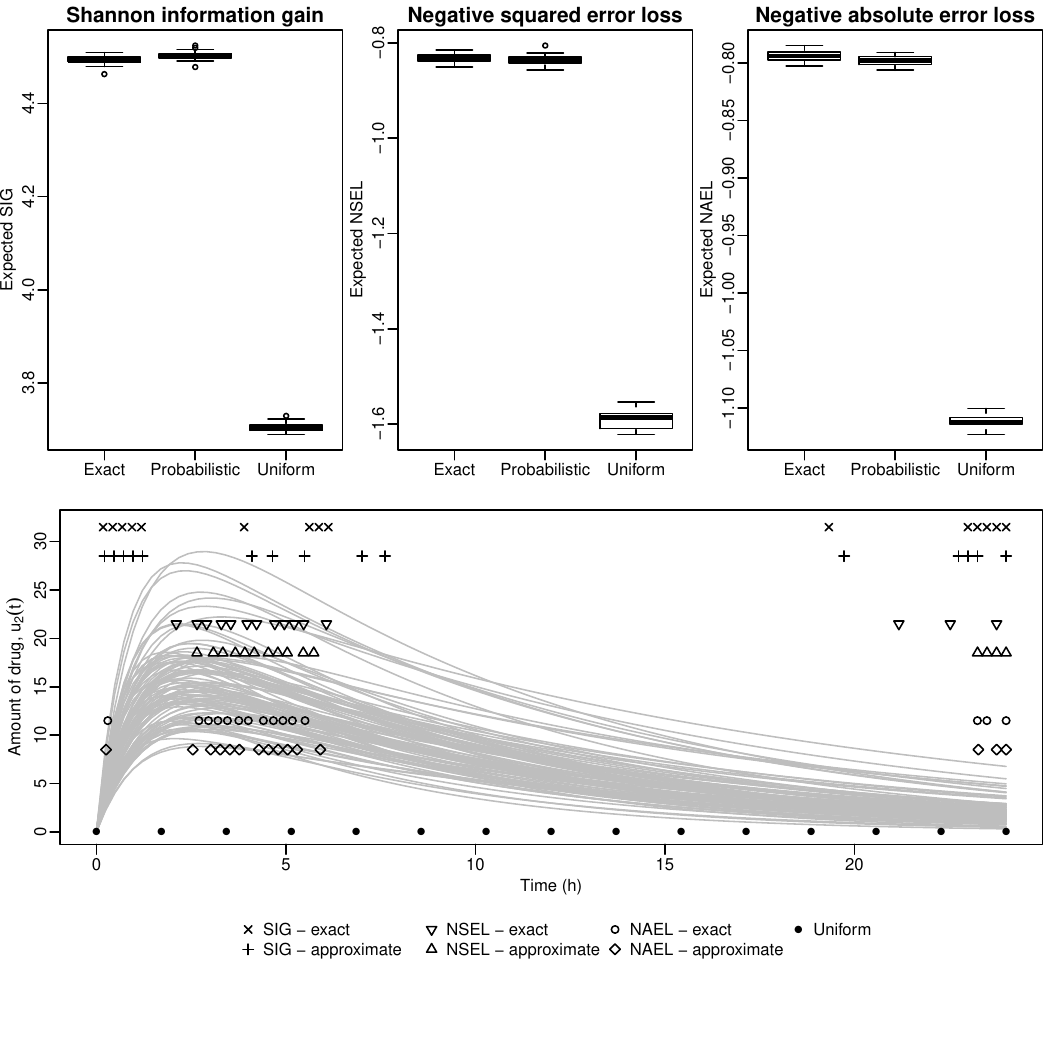}
\caption{\label{comp_plot} Results from the compartmental model in Section~\ref{EX_COMP}. Top row: boxplots of 20 evaluations of the Monte Carlo approximation to the expected utility for the uniform design and the optimal designs (for the exact and probabilistic solution) found under three different utility functions. Bottom plot: design points from each of the optimal designs and the uniform design, along with 100 draws from the exact solution, $u_2(t)$, giving the amount of drug at time $t$, for values drawn from the prior distribution of $\boldsymbol{\theta}$.}
\end{center}
\end{figure}

For each of the NSEL, NAEL and SIG utility functions from Section~\ref{sec:BOD_DTA}, we compare designs found under the exact and probabilistic solutions using ACE to a uniform design with $n=15$ equally-spaced time points in $[0,24]$ hours. Figure~\ref{comp_plot} presents boxplots of twenty evaluations of the Monte Carlo approximation to the expected utility for the uniform design and the optimal design found for each utility. There is negligible difference between the designs found under the exact and probabilistic solutions, and these designs are clearly superior to the uniform design. Figure~\ref{comp_plot} also gives the observation time points from each design being compared. The optimal designs appear to favor observation times near the peak of $u_2(t)$, at $t \approx 2.5$ hours, and then a series of observation times towards the end of the time interval. The optimal design under SIG has two distinct sets of points just before and after the maximum of $u_2(t)$, whereas the designs under NSEL and NAEL have just one set of points, generally occurring just after the peak response.

\subsection{FitzHugh-Nagumo equations \label{EX_FITZ}}
The FitzHugh-Nagumo equations (\citealp{F1961} and \citealp{N1962}) describe the behavior of spike potential in the giant axon of squid neurons:
$$\begin{array}{lll}
\dot{u}_1(t) & = & \theta_3\left[u_1(t) - u_1(t)^3/3 + u_2(t) \right]\,, \\
\dot{u}_2(t) & = & -\left[u_1(t) - \theta_1 + \theta_2 u_2(t)\right]/\theta_3\,, \\
\bu(0) & = & (-1,1)^{\T}\,,
\end{array}$$
where $u_1(t)$ is the voltage across the axon membrane, $u_2(t)$ is the recovery variable giving a summary of outward current, $\boldsymbol{\theta} = \left(\theta_1,\theta_1,\theta_3\right)^{\T}$, and $t \in [0,20]$ms. These equations cannot be solved analytically. 

We assume an experiment that measures the voltage, $y_l$, at time $t_l$, for $l=1,\dots,n$. Following \cite{RHCC2007}, $y_i \sim \mathrm{N}\left(u_1(t_i),\sigma^2 \right)$, independently, where $\sigma \sim \text{Uniform}[1/2,1]$. A priori, we assume $\theta_1,\theta_2 \sim \text{Uniform}[0,1]$ and $\theta_3 \sim \text{Uniform}[1,5]$.

\begin{figure}
\begin{center}
\includegraphics[scale=.9]{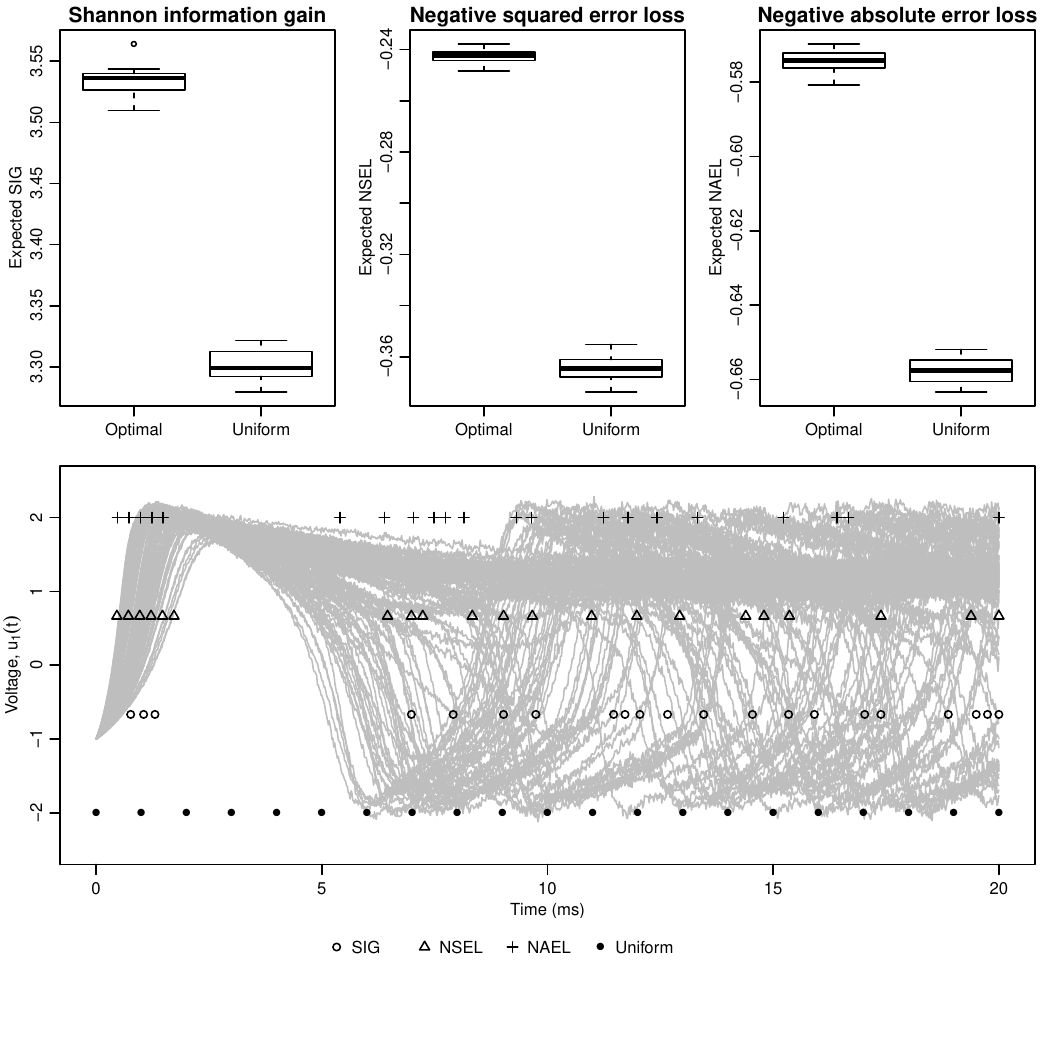}
\caption{\label{fitz_plot} Results from the FitzHugh-Nagumo equations in Section~\ref{EX_FITZ}. Top row: boxplots of 20 evaluations of the Monte Carlo approximation to the expected utility for the uniform design and the optimal designs found under three different utility functions. Bottom plot: design points from each of the optimal designs and the uniform design, along with 100 draws from the probabilistic solution, $u_1(t)$, giving the voltage at time $t$, for values drawn from the prior distribution of $\boldsymbol{\theta}$.}
\end{center}
\end{figure}

As noted by \cite{RHCC2007}, the solution to the FitzHugh-Nagumo equations can alternate between smooth evolution and sharp changes of direction. Hence, we employ uniform covariance~\eqref{eq:unifcov} for the probabilistic solution. The evaluation grid has size $N=200$.

The design consists of the $n=21$ observation times, $t_1,\dots,t_n$. Similarly to Section~\ref{EX_COMP}, we stipulate that the observation times must be at least 0.25ms apart, and find designs under the NSEL, NAEL and SIG utility functions. We compare these optimal designs to a uniform design with $n$ equally spaced points in $[0,20]$ms. Figure~\ref{fitz_plot} presents boxplots of twenty evaluations of the Monte Carlo approximation to the expected utility for the uniform design and the optimal designs found via ACE under each utility function. In each case, there is a clear improvement to be made over using the uniform design. Also shown in Figure~\ref{fitz_plot} are the four designs under comparison, along with realizations drawn from the solution $u_1(t)$. Both the NSEL and NAEL optimal designs have a substantial number of observations near the beginning of the experiment. Both these designs have around one-third of their observation times before 2.5ms; the SIG and uniform designs only make three observations before this time. A feature of all of the optimal designs is that they make no observations between about 2.5 and 6ms, where the voltage is expected to rapidly decrease. The remaining observation times are close to being evenly spaced. The initial phase of high frequency observations provides information about the steep increase in voltage for small $t$. The remaining observation times aid efficient parameter estimation, occurring within an interval within which different parameter values can produce very different model solutions.

\subsection{JAK-STAT mechanism \label{EX_JAKSTAT}}
\citet{CCGC2015}, and authors referenced therein, considered Bayesian inference for the JAK-STAT mechanism. A system of $s=4$ equations describes changes in the biochemical reaction states of STAT-5 transcription factors that occur in response to binding of the Erythropoietin hormone to cell surface receptors \citep{PD1997}:
\begin{equation*}
\begin{array}{lllll}
\dot{u}_1(t) & = &  - \theta_1u_1(t) \kappa(t) + 2\theta_4 u_4(t-\omega)\,, & \rdelim\}{4}{1em} & \multirow{4}{4cm}{$t \in [0,60]\text{ seconds}\,,$}\\
\dot{u}_2(t) & = & \theta_1u_1(t) \kappa(t) - \theta_2u_2(t)^2\,, & & \\
\dot{u}_3(t) & = & -\theta_3u_3(t) + \frac{1}{2}\theta_2 u_2(t)^2\,,& & \\
\dot{u}_4(t) & =  & \theta_3u_3(t) - \theta_4u_4(t - \omega)\,,& & \\
\bu(t) & =  & (u_{01},0,0,0)^{\T}\,, & & t \in [-\omega,0]\,,
\end{array}
\end{equation*}
with $u_{01}\ge 0$ unknown and $\kappa(t)$ an unknown forcing function. The transcription states return to the initial state after gene activation in the cell nucleus, modeled via the unknown time delay $\omega\ge 0$. This system is an example of a delay initial function problem.

\begin{figure}
\begin{center}
\includegraphics[scale=0.9]{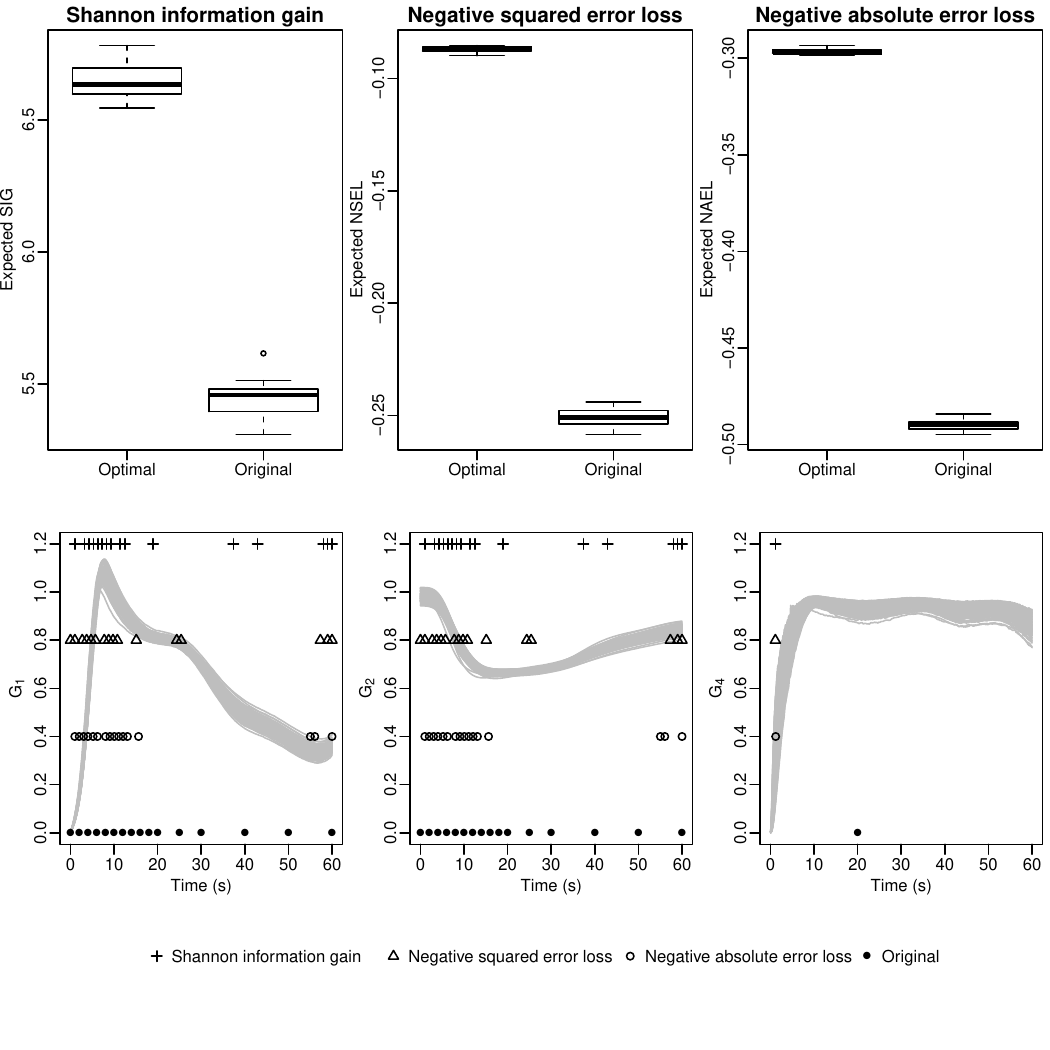}
\caption{\label{jakstat_plot} Results from the JAK-STAT example in Section~\ref{EX_JAKSTAT}. Top row: boxplots of 20 evaluations of the Monte Carlo approximation to the expected utility for the original design and the optimal designs found under three different utility functions. Bottom row: design points from each of the optimal designs and the original design at which noisy observations of $g_1$ (left), $g_2$ (center), $g_4$ (right) are made, along with 100 draws from $g_1$, $g_2$ and $g_4$, at time $t$, for values drawn from the prior distribution of $\boldsymbol{\theta}$, $u_{01}$ and $\omega$.}
\end{center}
\end{figure}

\citet{S2003} conducted an experiment that made measurements on the nonlinear transformation of the states given by
$$
g(\bu,\boldsymbol{\theta}) = \left(
\begin{array}{c}
\theta_5(u_2 + 2u_3)\\
\theta_6(u_1 + u_2 + 2u_3)\\
u_1\\
u_3/(u_2+u_3) \end{array}\right) 
= \left(
\begin{array}{c}
g_1(\bu,\btheta) \\
g_2(\bu,\btheta) \\
g_3(\bu,\btheta) \\
g_4(\bu,\btheta) \end{array}\right)\,.$$
The experiment made $n=16$ (noisy) observations on $g_1$ and $g_2$ at times $t_1,\ldots,t_{16}$, one observation on each of $g_3$ and $g_4$ at $t=0$ and $t=t^\star$, respectively. The design (choices of time points) used in the experiment reported by \citet{S2003} are given in Figure~\ref{jakstat_plot}.
The following statistical model is assumed
\begin{equation*}
\begin{split}
& (y_{1l},y_{2l})^{\T} \sim \mathrm{N}\left([g_1(\bu(t_l), \btheta),g_2(\bu(t_l), \btheta)]^{\T},\bA_l\right)\,, \\
& y_3 \sim \mathrm{N}\left(g_3(\bu(0),\boldsymbol{\theta}),\sigma_3^2 \right)\,,\quad y_4 \sim \mathrm{N}\left(g_4(\bu(t^\star),\boldsymbol{\theta}),\sigma_4^2 \right)\,,
\end{split}
\end{equation*}
independently, for $l=1,\dots,n$, where $\bA_l = \text{diag} \left\{\sigma_{1l}^2, \sigma_{2l}^2 \right\}$.

We design a follow-up experiment using information from this previous study, and choose values of $t_1,\ldots,t_n$ and $t^\star$ to maximize different expected utilities assuming, for simplicity, a single observation of $y_3$ will also be made at $t=0$ (as in the original experiment). We use the posterior distributions from \citet{CCGC2015} as priors for $\btheta$, $\omega$ and $u_{01}$. These authors assumed the variance parameters were fixed. Instead, we assume $\sigma^2_{1l} = \sigma^2_1$, $\sigma^2_{2l} = \sigma^2_2$, for all $l=1,\dots,n$, and $\sigma_1,\sigma_2 \sim \mathrm{Uniform}[0,0.1]$, $\sigma_3 \sim \mathrm{Uniform}[0,20]$ and $\sigma_4 \sim \mathrm{Uniform}[0,0.1]$. These prior distributions are consistent with the experimentally determined values used for previous analyses (see \citealp{R2009}). The forcing function $\kappa(t)$ is assumed unknown but has been measured at 16 time points. We follow \citet{CCGC2015} and assume these measurements are made without error and interpolate with a Gaussian process to allow a probabilistic prediction of $\kappa(t)$ for any $t\in [0,60]$.

The nature of the delay initial function problem introduces an added complexity to our implementation of the probabilistic solution. At the end of step~\ref{alg:update:iter} of Algorithm~\ref{alg:update}, we compute $\beff_{r+1} = \beff(\bu(\tau_{r+1}), \tau_{r+1},\boldsymbol{\theta}_b)$. For this example, to compute $\beff_{r+1}$, we require $u_4(\tau_{r+1}-\omega_b)$, where $\omega_b$ is a value generated from the prior distribution of $\omega$. If $\tau_{r+1} - \omega_b \le 0$, then $u_4(\tau_{r+1}-\omega_b)=0$ as specified by the initial conditions of the system of equations. For $\tau_{r+1} - \omega_b > 0$, the conditional distribution of $u_4(\tau_{r+1}-\omega_b)$ can be derived in the probabilistic solution of \cite{CCGC2015} and a value for $u_4$ generated. However, this will be computationally expensive to incorporate in the implementation of the probabilistic solution described in Section~\ref{sec:extode} and would prevent the precomputation in Algorithm~\ref{alg:precompute}. Hence, if $\tau_{r+1} - \omega_b >0$, we replace $u_4(\tau_{r+1} - \omega_b)$ by $u_4(\tau_{\bar{r}})$, where $\bar{r} = \mathrm{arg} \min_{r'=1,\dots,r+1} |\tau_{r+1} - \omega_b - \tau_{r'}|$, i.e. from the series of $u_4(\tau_1),\dots,u_4(\tau_{r+1})$ values generated in step~\ref{alg:update:iter} thus far, we choose the value for the time $\tau_{\bar{r}}$ that is closest in absolute value to $\tau_{r+1} - \omega_i$.

We employ uniform covariance~\eqref{eq:unifcov} as the time delay can cause discontinuities in the derivative, as noted by \citet{CCGC2015}. The evaluation grid, $\boldsymbol{\tau}$, has size $N=500$, and the auxiliary parameters are set to $\lambda = 0.085$ and $\alpha = 8000$, consistent with the posterior distribution from the original analysis.

We use the methodology from Section~\ref{sec:extode} and the ACE algorithm to find designs that maximize each of the NSEL, NAEL and SIG utilities. We compare these designs to the original design used by \cite{S2003}. As in the previous examples, we introduce the constraint that the observation times need to be at least 1 second apart, a requirement also satisfied by the original experiment. Figure~\ref{jakstat_plot} presents boxplots of twenty evaluations of the Monte Carlo approximation to the expected utility for the original design and the optimal designs found under each utility function. Once again, in each case, the optimal designs are considerably more efficient. Also shown in Figure~\ref{jakstat_plot} are the four designs under comparison. The optimal designs favor having a dense set of points early in the observation window, and then a smaller set of times near the end of the experiment. This is especially true for the designs under NSEL and NAEL where 75\% of the observation times occur before $t=15$ seconds, compared to about 60\% for SIG design and 50\% for the original design. Early observation times provide information about the peak in $g_1$ and the sharp decrease in $g_2$ at about 10 seconds. For the single observation time, $t^*$, on $g_4$, the optimal designs clearly favor making a very early observation. Note that $t^*$ for each of the optimal designs is between 1 and 2 seconds.


\section{Application: transport of serine across human placenta \label{sec:EX_PLAC}}
We now use the methodology in Section~\ref{sec:ODEdesign} to redesign the experiment for the human placenta study introduced in Section~\ref{sec:INTRO}. The experimental protocol specifies fixed initial amounts of radioactive serine interior ($u_{01}$) and exterior ($x_1$) to the placenta ($0$ and $7.5\mu$l, respectively). The original design proposed by the experimenters used $n=7$ placentas (runs) with differing amounts of non-radioactive serine interior ($u_{01}$) and exterior ($x_2$) to the placenta, see Table~\ref{plac_tab}. Noisy observations on the amount of interior radioactive serine ($u_1$) were made at eight times, common to each of the seven placentas. The experimenters expected greater variability in the concentration of interior radioactive serine near the start of the experiment, before convergence to an equilibrium. Therefore, they choose a design containing a large number of early time points. We broadly follow this protocol, but find optimal designs using $n=2,\ldots,7$ placentas with each having $n_t=8$ observations taken at common times, $t_1,\ldots,t_8$, chosen from across the interval $[0, 600]$. 

A hierarchical statistical model is assumed for the observed responses:
\begin{equation*}
y_{jl} = u_1(t_l; \bx_j, \boldsymbol{\theta}_j) + \varepsilon_{jl}\,,\qquad\text{for } j=1,\ldots,n;\, l=1,\ldots,n_t\,,
\label{plac_model}
\end{equation*}
where $\bx_j = (x_1,x_{2j})^{\T}$, $\varepsilon_{jl}$ are independent and identically normally distributed with constant variance $\sigma^2$, and $\boldsymbol{\theta}_j$ holds the $p=4$ subject-specific parameters for the $j$th placenta with elements assumed to follow independent uniform distributions
$$\theta_{ji} \sim \mathrm{U}\left[ \theta_i\left(1-c_i\right), \theta_i\left(1+c_i\right)\right]\,,\qquad c_i>0\,,\qquad i=1,\ldots,p\,.$$
The goal of the experiment is estimation of the population physical parameters $\btheta = (\theta_1,\ldots,\theta_p)^{\T}$.

A priori, we assume $c_i \sim \text{Uniform}\left[0,0.05\right]$ and $\theta_i \sim \mathrm{Tri}[a_i,b_i]$, where $\text{Tri}[a,b]$ denotes the symmetric triangle distribution on the interval $[a,b]$. Reflecting prior knowledge from previous experiments, we set $a_1=a_3=a_4=80$, $b_1=b_3=b_4=120$, $a_2=0.02$, $b_2= 0.08$ and we assume $\sigma^2 \sim \mathrm{U}[0,1]$ for the response variance. 

We expect the solution to system of equations~\eqref{eq:odes2b} to be smooth, and so use squared exponential covariance~\eqref{eq:sqexpcov} for the probabilistic solution. The evaluation grid, $\boldsymbol{\tau}$, has size $N=601$ and we set auxiliary correlation parameter $\alpha = 10N$.

Specifying a design corresponds to specifying the $n$ experimental conditions $x_{21},\dots,x_{2n}$, initial values $u_{021},\dots,u_{02n}$, and the common $n_t = 8$ observation times $t_1,\dots,t_{n_t}$. Hence for $n=2,\ldots,7$, the design space has between 12 and 22 dimensions. As for the examples in Section~\ref{sec:EXAMPLES}, we impose a constraint on the observation times and specify that they must be at least 5 seconds apart.

We find designs for the NSEL, NAEL, 0-1 estimation and 0-1 model selection utility functions defined in Section~\ref{sec:BOD_DTA}. For the 0-1 estimation utility, we set $\bdelta = (5, 5, 0.01, 5)^\T$; for utility $\phi(\btheta,\by,\bd)$ to equal 1, the posterior mean for $\btheta$ must lie in the box set $\prod_{i=1}^4[\theta_i-\delta_i, \theta_i + \delta_i]$, which contains 0.5\% of the volume of the prior support.  For the model selection utility, we suppose interest is in determining if the reaction rates are equal, i.e. does $\theta_3 = \theta_4$? To answer this question, we define two models: $m_1$ (where $\theta_3=\theta_4$) and $m_2$ (where $\theta_3 \ne \theta_4$). 

\begin{figure}
\begin{center}
\includegraphics[scale=0.9]{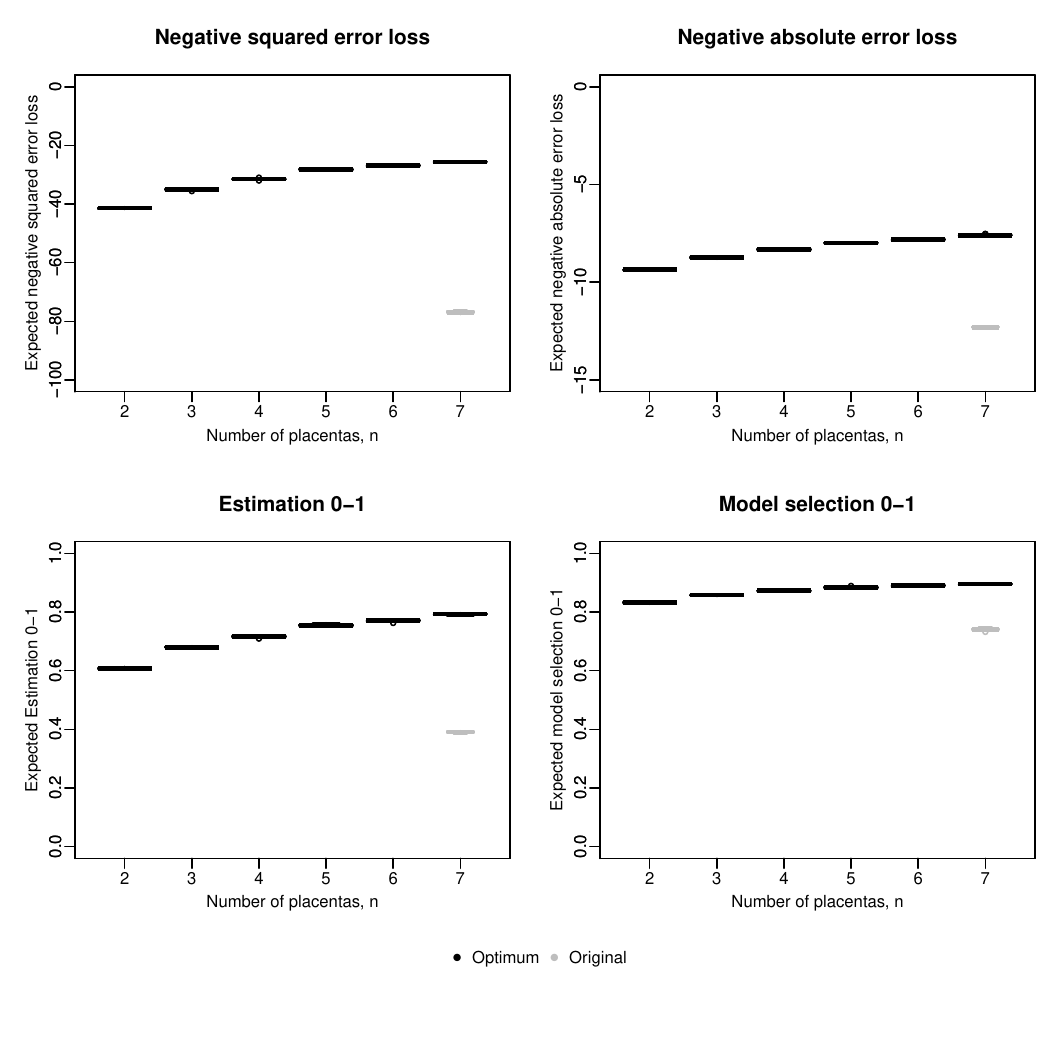}
\caption{\label{plac_plot0} Results from the placenta example in Section~\ref{sec:EX_PLAC}. Boxplots of 20 evaluations of the Monte Carlo approximation to the expected utility for the original design and the optimal designs found under four different utility functions for $n=2,\ldots,7$.}
\end{center}
\end{figure}

Figure~\ref{plac_plot0} presents boxplots of twenty evaluations of the Monte Carlo approximation to the expected utility for the optimal design found under each utility function for $n= 2,\dots,7$. We also present boxplots of the performance of the original design with $n=7$. Unsurprisingly, the expected utility increases with $n$, and the optimal designs are clearly superior to the original design. For each utility function, the optimal design with $n=2$ outperforms the original design with $n=7$ placentas, with substantial differences in expected utility.     

\begin{table}
\begin{center}
\caption{Treatments from the optimal and original designs with $n=7$ runs for the placenta example in Section~\ref{sec:EX_PLAC}: initial concentrations (to nearest integer) of interior ($u_{02} = u_2(0)$)  and exterior ($x_2$) non-radioactive serine for each run (placenta).\label{plac_tab}}
\medskip
\begin{tabular}{l|rr|rr|rr|rr|rr} \hline
 & \multicolumn{2}{|c|}{NSEL} & \multicolumn{2}{|c|}{NAEL} & \multicolumn{2}{|c|}{Est01$^\dagger$} & \multicolumn{2}{|c|}{MS01$^\star$} & \multicolumn{2}{c}{Original} \\ \hline
Placenta & $x_2$ & $u_{02}$ & $x_2$ & $u_{02}$ & $x_2$ & $u_{02}$ & $x_2$ & $u_{02}$ & $x_2$ & $u_{02}$ \\ \hline
1 &  0 & 0 			& 0 & 0 		& 0 & 0	 	& 0 & 0 		& 0 & 0 \\
2 &  0 & 38 		& 0 & 0 		& 0 & 0	 	& 0 & 0 		& 250 & 0 \\
3 &  0 & 50 		& 0 & 50 		& 0 & 56	 	& 0 & 0 		& 250 & 250 \\
4 &  0 & 68 		& 0 & 67 		& 0 & 58	 	& 0 & 0 		& 250 & 1000 \\
5  & 182 & 1000 	& 160 & 1000 	& 177 & 1000	& 0 & 38	 	& 1000 & 0 \\
6  & 185 & 1000 	& 175 & 1000 	& 196 & 1000	& 0 & 41	 	& 1000 & 250 \\
7  & 206 & 1000 	& 211 & 1000 	& 210 & 1000	& 115 & 62	 & 1000 & 1000 \\ 
\hline 
\end{tabular}
\end{center}
\vspace*{-2ex}
\footnotesize
\hspace*{10ex} $^\dagger$ 0-1 estimation utility; $^\star$ 0-1 model selection utility 
\end{table}

\begin{figure}
\begin{center}
\includegraphics[scale=.9]{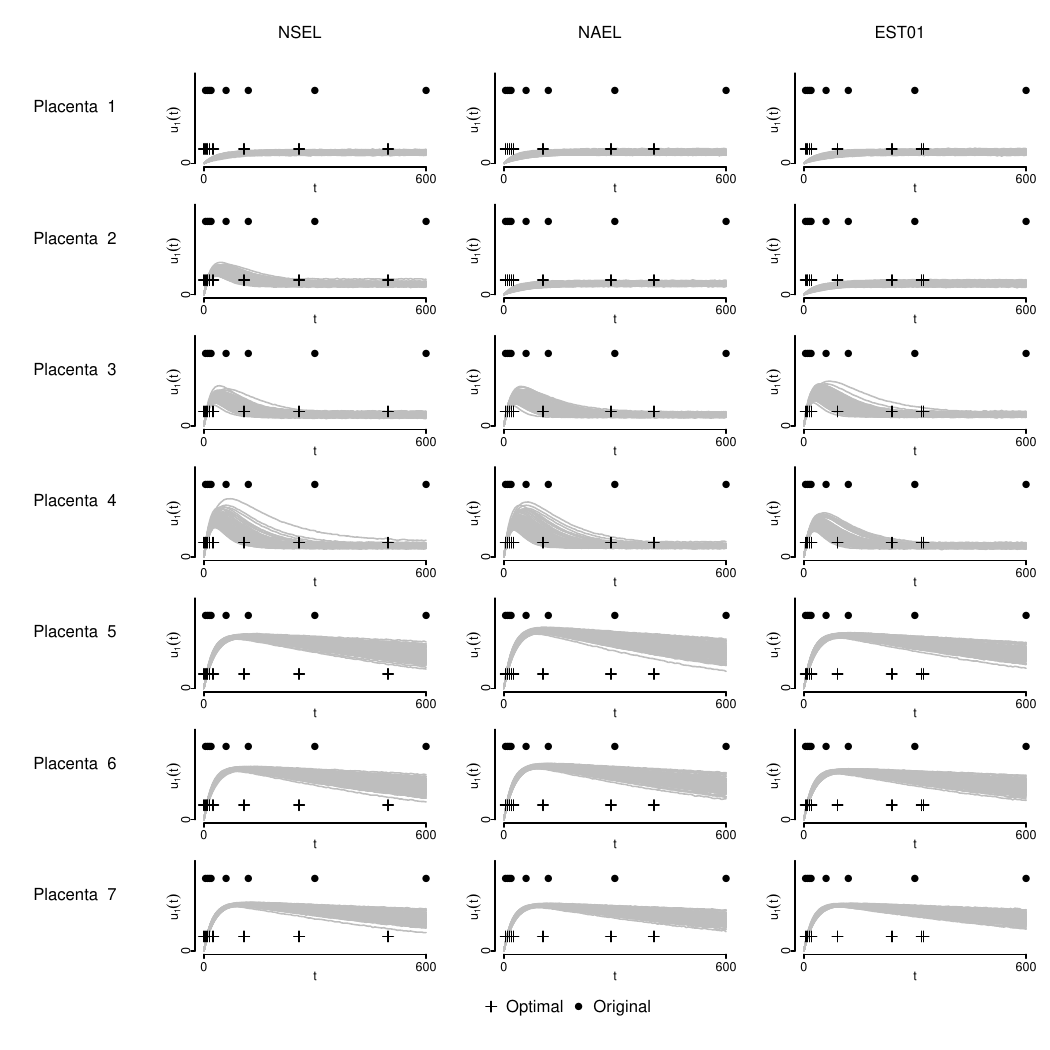}
\caption{\label{plac_plot} Results from the designs found under SIG, NSEL and NAEL utilities with $n=7$ placentas in Section~\ref{sec:EX_PLAC}. Displayed are 100 draws from solution $u_1(t)$ plotted against $t$ for values drawn from the the prior distribution of $\boldsymbol{\theta}$, for each of the $n=7$ placentas and treatments given in Table~\ref{plac_tab}.}
\end{center}
\end{figure}

Table~\ref{plac_tab} gives the treatments for each design found for $n=7$. Figure~\ref{plac_plot} shows the observation times for the optimal designs under NSEL, NAEL and 0-1 estimation utilities, along with realizations from the solution to $u_1(t)$, for each run of each design. The designs under NSEL and NAEL utilities have similar treatments and observation times. The initial concentrations in Table~\ref{plac_tab} lead to three distinct profiles of $u_1(t)$ (labeled placentas 1 and 2; 3 and 4; 5, 6 and 7; note though that the placentas are exchangeble). The profile for placentas 1 and 2 has a slow steady increase in $u_1(t)$ with respect to $t$. Placentas 3 and 4 have a steep initial increase and subsequent decrease in $u_1(t)$ with respect to $t$. Finally, placentas 5 to 7 have a steep initial increase in $u_1(t)$ with respect to $t$ followed by a slow decrease. The optimal observation times are predominantly at the beginning of the observation window, where $u_1(t)$ is changing most quickly. The designs under the 0-1 estimation utility are also similar, except a non-zero amount (35 $\mu$l) of the initial interior non-radioactive serine is applied to placenta 2.


\begin{figure}
\begin{center}
\includegraphics[scale=.9]{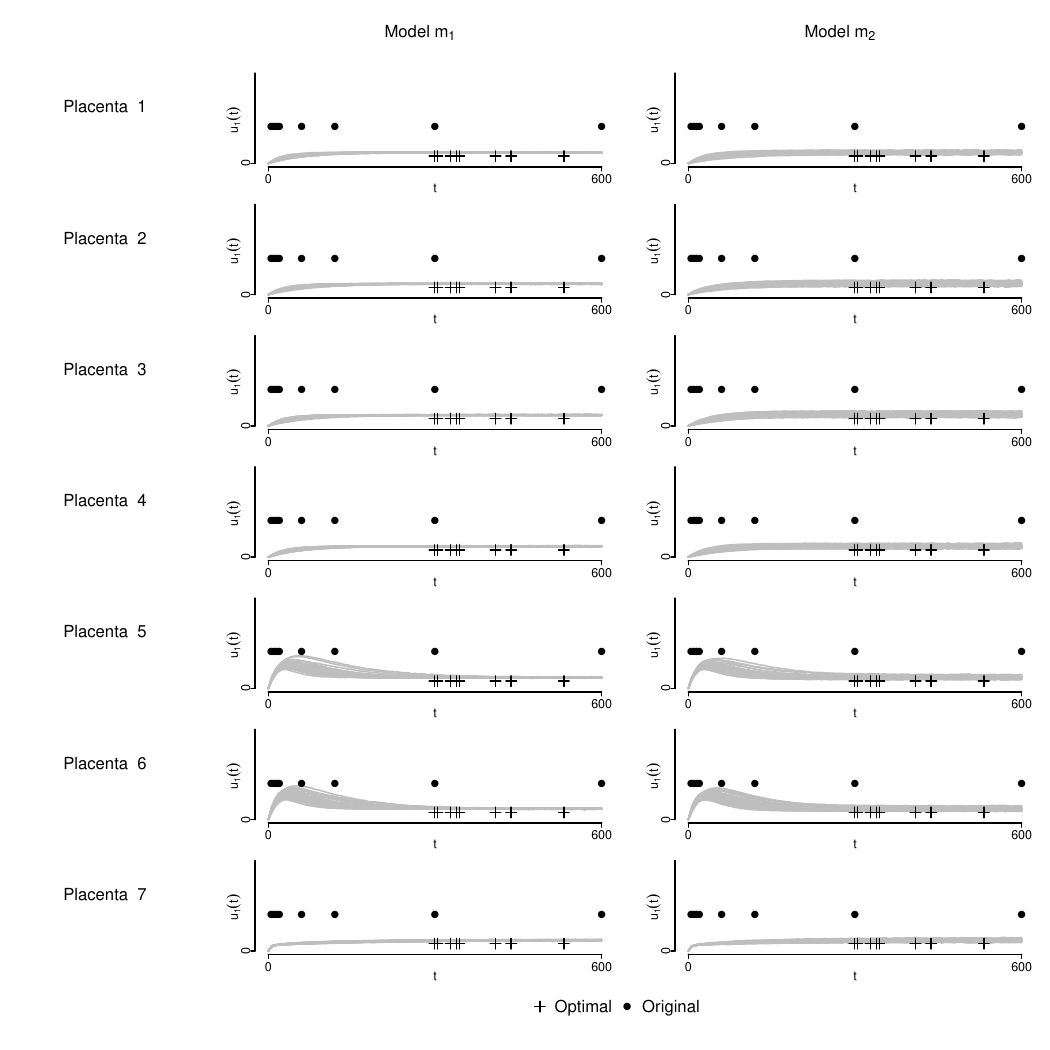}
\caption{\label{plac_plot2} Results from the designs under the 0-1 model selection loss with $n=7$ placentas in Section~\ref{sec:EX_PLAC}. Displayed are 100 draws from solution $u_1(t)$ under model $m_1$ ($\theta_3=\theta_4$) and model $m_2$ ($\theta_3\ne\theta_4$) plotted against $t$ for values drawn from the the prior distribution of $\boldsymbol{\theta}$, for each of the $n=7$ placentas and treatments given in Table~\ref{plac_tab}.}
\end{center}
\end{figure}

Figure~\ref{plac_plot2} shows the designs from the 0-1 model selection utility, along with realizations of the solutions $u_1(t)$ under models $m_1$ and $m_2$. The treatments for the optimal design under the 0-1 utility result in two distinct profiles of $u_1(t)$. For placentas 1--5, $u_1(t)$ has a slow steady increase in $u_1(t)$ with respect to $t$. Placentas 6 and 7 have a steep initial increase and subsequent decrease in $u_1(t)$ with respect to $t$. Unlike the other optimal designs, the observation times are predominantly towards the end of the observation window. The $u_1(t)$ profiles are similar under both models, with the most substantial differences occurring in the inter-profile variability towards the middle of the time interval. This region is where the majority of observation times are located.

The original design proposed by the experimenters had an unequal spacing of observation times across the entire interval $[0,600]$. There are more observations taken near the start of the interval, and the time points are not dissimilar to those in the optimal designs under NSEL, NAEL and 0-1 estimation. However, the original design has treatments that are very different from any of the optimal designs, with an almost factorial structure and some treatments with high values of $x_2$ (exterior initial concentration of non-radioactive serine). None of the optimal designs include treatments with high $x_2$, demonstrating how it is often difficult to predict by intuition the treatments in a Bayesian optimal design for a complicated nonlinear model. In addition, the designs for point estimation (under NSEL, NAEL and 0-1 estimation utilities) are quite different to the design for model selection.

\section{Concluding Remarks}\label{sec:disc}
This paper introduces and demonstrates the first practical methodology for Bayesian optimal design of experiments for statistically nonlinear models formed from the solution to intractable ordinary differential equations. The work is motivated by a challenging design problem from the biological sciences, which we address through a combination of probabilistic solutions to the equations, simulation-based approximation to expected utilities and optimization via smoothing and cyclic descent. Our novel adjustments to the \citet{CCGC2015} probabilistic algorithm are key to providing a computationally efficient solution to the optimal design problem. Through demonstration on a number of examples, including the motivating experiment on serine transport across placental membranes, we show the efficiency gains that can be made by use of optimal designs over obvious, and proposed, alternatives. We also show how it is often not possible to ``second guess'' via intuition the solutions to optimal problems for nonlinear models.  

We have adopted the nested integration and optimization methods from \citet{OW2015} to find optimal designs for the differential equation models in this paper (namely the ACE algorithm). The Markov chain simulation schemes of \citet{M1999} and \citet{MSI2004}, among other authors, would be an interesting alternative approach. Extension and application of such methods to the problems in the current paper is an area for future research.

One key issue not addressed is model discrepancy (see, e.g., \citealp{KOH2001} and \citealp{plumlee2017}); the systematic mis-match between the true physical process and the solution to the ordinary differential equations. Not taking account of this error can lead to significant bias in posterior estimates of the physical parameters \citep{BOH2014}. Future work will focus on Bayesian optimal design for physical models subject to model discrepancy.

Some limited insight into the impact of model mis-match can be gained from a simple extension to the compartmental model in Section~\ref{EX_COMP}. For the purpose of finding designs, we assume the model
\begin{equation}\label{eq:wrong}
u_2(t)  = \frac{D}{0.9 \theta_3} \left( \exp \left( - \theta_1 t\right) - \exp \left( - \theta_2 t\right)\right)\,.
\end{equation}
That is, we simplify~\eqref{eq:compsol2} by setting $\theta_1$ and $\theta_2$ equal to their prior means in the fraction that multiplies the exponential term. We still assume the exponential depends on unknown $\theta_1, \theta_2$, and assume the same prior distributions for all parameters as in Section~\ref{EX_COMP}. 

\begin{figure}
\begin{center}
\begin{tabular}{ccc}
\includegraphics[scale=0.29]{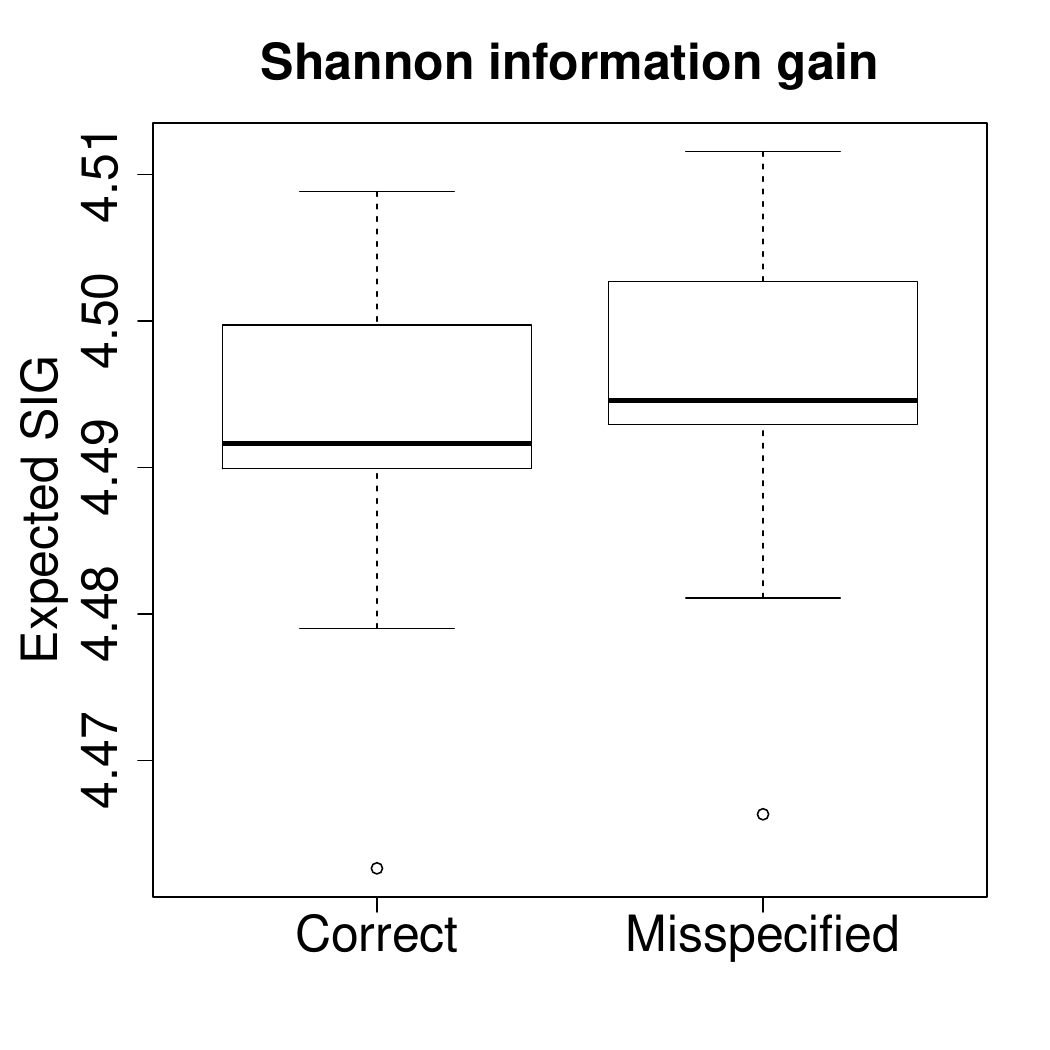} & \includegraphics[scale=0.29]{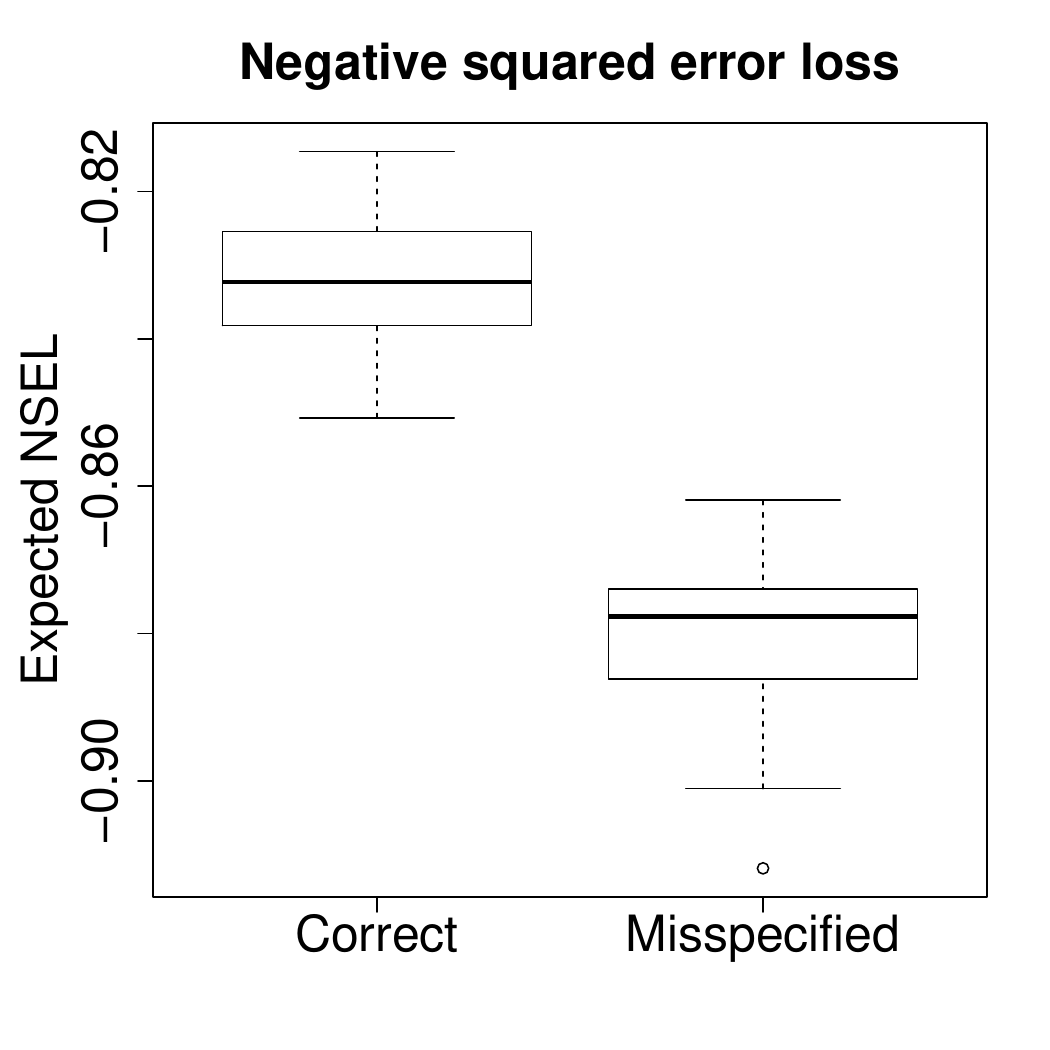} & \includegraphics[scale=0.29]{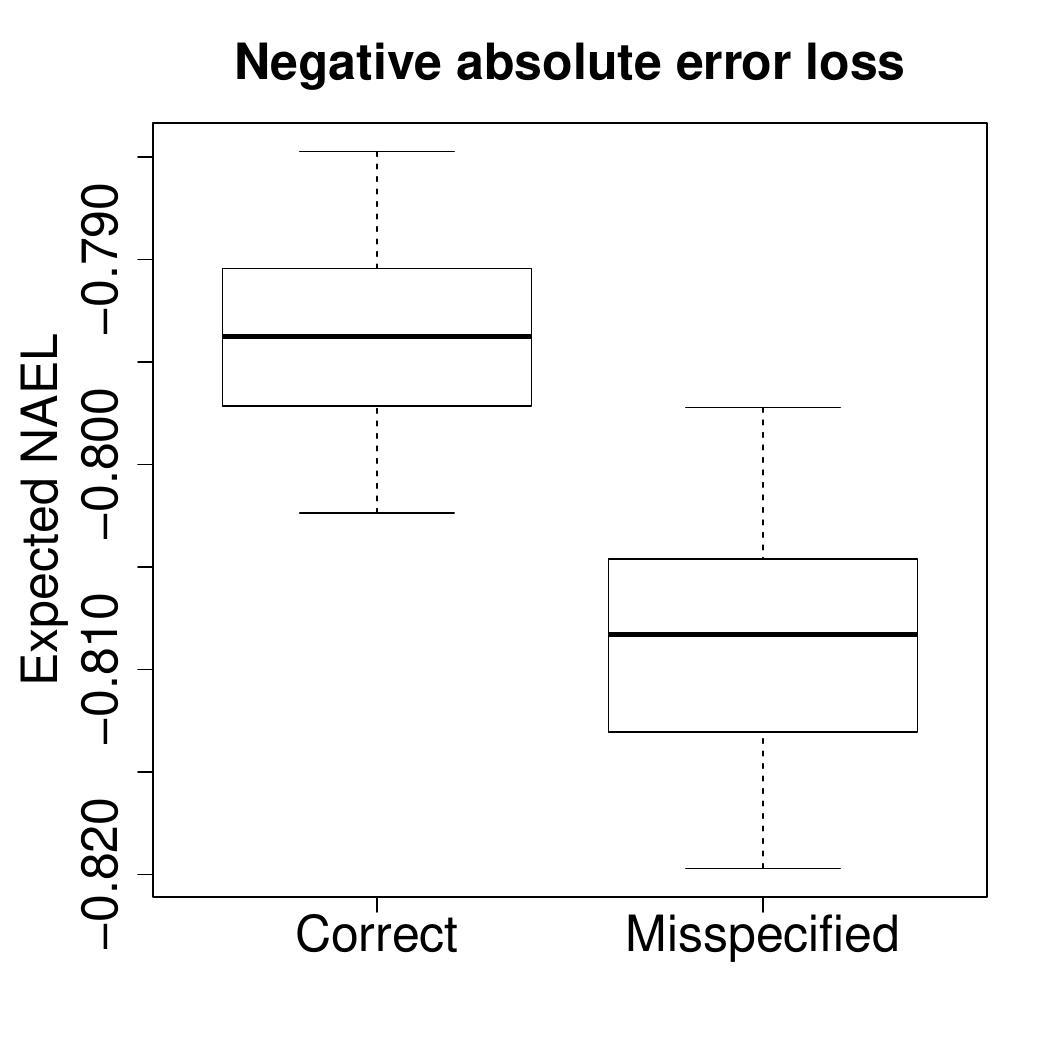}
\end{tabular}
\caption{\label{fig:wrong} Results from the misspecified model example in Section~\ref{sec:disc}. Boxplots of 20 evaluations of the Monte Carlo approximation to the expected utility for the optimal designs found under the correct model~\eqref{eq:compsol2} and the misspecified model~\eqref{eq:wrong} under the SIG, NSEL and NAEL utility functions. In each case, the correct model is assumed for evaluating the expected utility.}
\end{center}
\end{figure}

To assess the impact of model mis-match, we find optimal designs under the SIG, NSEL and NAEL utilities assuming the misspecified model~\eqref{eq:wrong}. We then assess these designs under the correct, more complex, model~\eqref{eq:compsol2} and compare them to designs found under the correct model by evaluating the approximate expected utility under the correct model, see Figure~\ref{fig:wrong}. Differences in approximate expected utility between the designs found under the correct and misspecified models are comfortably within Monte Carlo error for SIG. However, assuming a misspecified model under the NSEL utility results in a loss of expected utility of around 6\%; the differences are somewhat less for NAEL but still larger than Monte Carlo error. Clearly, the reduction in expected utility from assuming a misspecified model will depend on the models under consideration, the difference between the models and the choice of prior distributions, in addition to the choice of utility function. This is an important area for future research.       

\section*{Acknowledgements}
We thank Rohan Lewis, Bram Sengers and Nont Panitchob from Medicine, Life Sciences and Bioengineering at the University of Southampton for providing details of the placenta experiment. We are grateful to two anonymous reviewers whose comments improved the paper.

\section*{Funding}
The second author was supported by Fellowship EP/J018317/1 from the United Kingdom Engineering and Physical Sciences Research Council.

\appendix

\section{The ACE algorithm} \label{app:ACEalg}
In Algorithm~\ref{alg:ace}, we outline the basic approximate coordinate exchange algorithm. For full details, see \citet{OW2015}.

Let $M$ be the total number of coordinates (values taken by each variable in each run) of the design. That is, $M$ is the dimension of the design. In step~\ref{ace:p}, the probability of the suggested design having higher expected utility is calculated, a posteriori to two independent Monte Carlo samples from the joint distributions of the data and parameters conditional on the current and suggested designs. Calculation of this probability assumes the utility evaluations are well described by a normal distribution. In the case of 0-1 utilities, a similar test based on a Bernoulli likelihood and Beta prior is applied (see \citealp{OMD2018} for details). Convergence in step~\ref{ace:conv} is assessed informally using trace plots of the evaluations of either $\bar{\phi}^*$, if the proposed design was accepted, or $\bar{\phi}^C$, otherwise, from step~\ref{ace:p}.

The ACE algorithm should be started from multiple different starting designs $\bd^0$. From the resulting designs, the one with the lowest value of $\hat{\Phi}(\bd)$ should be returned.

\spacingset{1}
\begin{algorithm}
\DontPrintSemicolon
\nl Choose an initial design $\bd^0 = \left(d_1^0,\dots,d_M^0\right)^{\T}$ and set the current design to be $\bd^C = \left(d_1^C,\dots,d_M^C\right)^{\T}=\bd^0$\;
\nl \label{ace:loop}\For{$i = 1:M$}{
\nl Generate a one-dimensional space-filling design $\zeta_{i} = \left\{d_{i}^1,\ldots,d_{i}^R\right\}$ in $\mathcal{D}_i\subset\mathbb{R}$, the set of possible values for the $i$th coordinate\;
\nl Let $\bd^C(d_{i}^r)$ equal $\bd^C$ with $i$th coordinate replaced by $d_{i}^r$\;
\For{$r = 1:R$}{Evaluate $\hat{\Phi}(\bd^C(d_{i}^r))$, the approximation to the expected utility, i.e. equation~\eqref{eq:mc}
}
\nl Fit a Gaussian process emulator $\tilde{\Phi}(d)$ using ``data'' $\left\{d_{i}^r, \hat{\Phi}(\bd^C(d_{i}^r))\right\}_{r=1}^R$\;
\nl Set $\bd^\star = \left(d_1^C,\ldots,d^C_{i-1},\tilde{d},d^C_{i+1},\ldots,d^C_M\right)^{\T}$, where $\tilde{d}\in \argmax_{d\in\mathcal{D}_i}\tilde{\Phi}(d)$\;
\For{$j=1:B$}{
\nl Generate $\left[(\by_j^C)^\T,(\btheta_j^C)^\T\right]^\T\sim\pi(\btheta,\by | \bd^C)$ and $\left[(\by_j^\star)^\T,(\btheta_j^\star)^\T\right]^\T\sim\pi(\btheta,\by | \bd^\star)$\;
\nl Set $\phi_j^C = \phi(\btheta_j,\by_j,\bd^C)$ and $\phi_j^\star = \phi(\btheta_j,\by_j,\bd^\star)$\;
}
\nl \label{ace:p} Calculate
$$
p^* = 1 - F_{t,2B-2}\left(-\frac{\sum_{i=j}^B \phi_i^C - \sum_{i=1}^B \phi_j^*}{\sqrt{2B \hat{v}}}\right)\,,
$$
where $F_{t, a}(\cdot)$ is the distribution function of the $t$-distribution with $a$ degrees of freedom,
$$\hat{v} = \frac{\sum_{j=1}^B (\phi_j^C - \bar{\phi}^C)^2 + \sum_{j=1}^B (\phi_j^* - \bar{\phi}^*)^2}{2B - 2}\,,$$
and $\bar{\phi}^C = \sum_{j=1}^B\phi^C_j/B$ and $\bar{\phi}^\star= \sum_{j=1}^B\phi^\star_j/B$\;
\nl Set $\mathbf{d}^C = \mathbf{d}^\star$ with probability $p^*$\;
}
\nl \label{ace:conv} Return to step~\ref{ace:loop} until convergence.
\caption{\label{alg:ace}The approximate coordinate exchange (ACE) algorithm.}
\end{algorithm}
\spacingset{1.45}

\newpage

\bibliographystyle{asa}
\bibliography{mybib}

\end{document}